\documentclass[peerreview]{IEEEtran}
\usepackage{cite} 
\usepackage{url} 
\usepackage[utf8]{inputenc} 
\usepackage{booktabs} 
\usepackage{graphicx}
\usepackage{amsmath}
\usepackage{xcolor} 
\usepackage{subcaption}

\begin{document}
\title{Time Slicing of Neutrino Fluxes in Oscillation Experiments at Fermilab}


\author{Sudeshna Ganguly, Katsuya Yonehara\\
Chandrashekhara M Bhat, A. Kent Triplett\\
Robert Ainsworth\\
Fermilab\\
Clara Hinds\\
University of Chicago\\
Maan Abdelhamid\\
University of Illinois, Urbana-Champaign\\
}
\date{09/28/2024}

\maketitle
\tableofcontents
\listoffigures
\listoftables

\IEEEpeerreviewmaketitle
\begin{abstract}
Long and short baseline neutrino oscillation experiments, such as DUNE, ANNIE, SBND, demand high precision in reducing systematic errors, particularly those related to neutrino-nucleus interaction cross-sections. The stroboscopic approach offers a method to capture distinct neutrino energy spectra, aiding in the separation of flux and cross-section uncertainties. This report outlines the creation of short proton bunch lengths and the transport of a narrow beam down the Booster Neutrino Beamline, essential steps for the successful implementation of the stroboscopic approach. 
\end{abstract}

\section{Introduction}
The next generation of long and short baseline neutrino experiments aims to increase proton beam power to multi-MW level and make use of massive detectors to overcome the limitation of event statistics. The Deep Underground Neutrino Experiment (DUNE) experiment at the Long Baseline Neutrino Facility (LBNF) \cite{lbnfbeamline} will test the three neutrino flavor paradigm and directly search for CP violation by studying oscillation signatures in the high intensity $\nu_{\mu}$ (anti-$\nu_{\mu}$) beam to $\nu_e$ (anti-$\nu_e$) measured over a long baseline. 
Enhanced knowledge of the neutrino energy spectrum is crucial for improving event reconstruction and reducing systematics. One of the most significant sources of systematic uncertainty arises from neutrino-nucleus interaction cross sections. A better understanding of these cross sections is essential for improving the precision of DUNE's oscillation measurements and advancing neutrino physics.\\
The stroboscopic approach \cite{angelico2019neutrino} introduces a powerful method for capturing distinct neutrino energy spectra using on-axis detectors, providing additional leverage to control both flux and cross-section related systematics. By generating short proton bunch lengths (O(100 , \text{ps})), implementing fast detector timing, and synchronizing with proton bunches at the target, this technique allows for more precise neutrino energy estimation.
When applied to long- and short-baseline experiments such as the DUNE or SBND, the stroboscopic approach could significantly improve the precision of neutrino energy measurements and aid in the search for new physics.\\
To apply this approach, narrow proton beams are essential.
This report focuses on creating narrow proton beams for the Booster Neutrino Beam (BNB) target and characterizing beam losses, with potential applications to both long and short baseline neutrino oscillation experiments.\\

\section{Added Value of Stroboscopic Approach to Neutrino Oscillation Experiments}
\begin{figure}[htpb!]
  \begin{center}\setlength{\unitlength}{0.2cm}
   \includegraphics[width=0.45\textwidth]{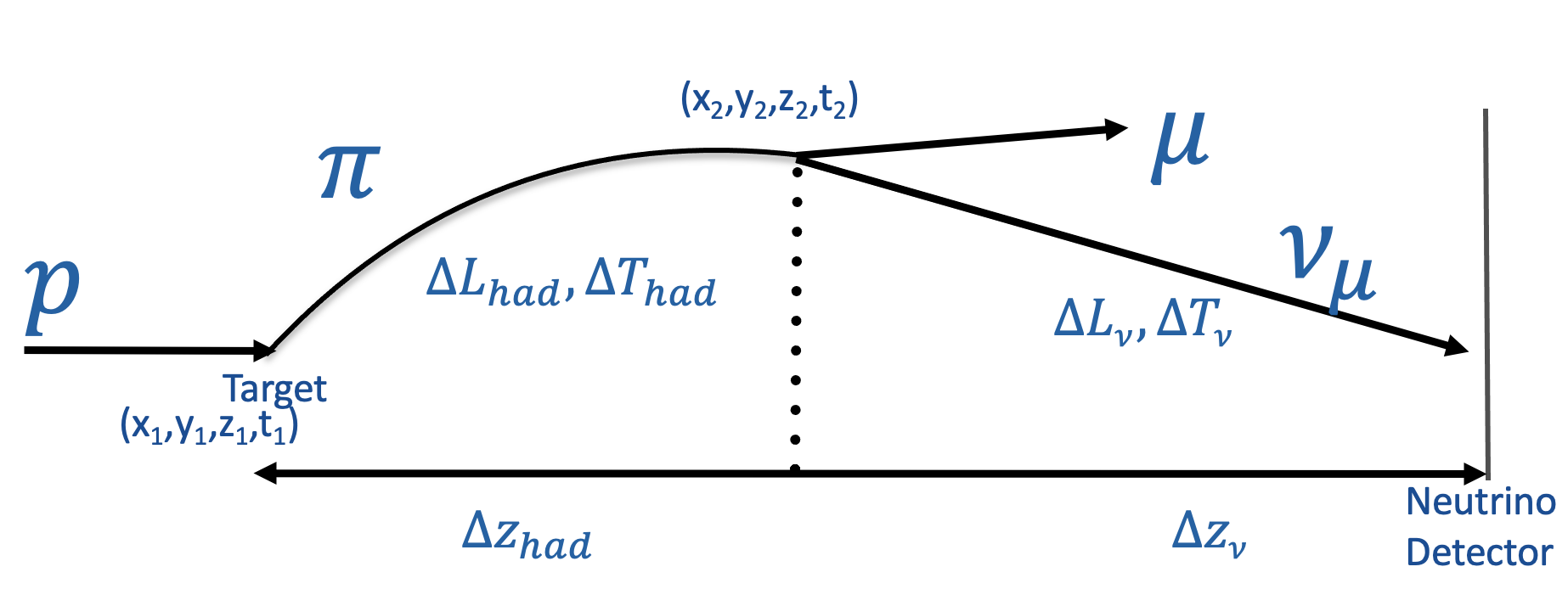}
\caption{Diagram of a proton impinging on the target and producing hadrons (pions), which then decay into muons and neutrinos.}\label{fig:deltat_calc}
   \end{center}
  \end{figure}

The stroboscopic approach leverages the correlation between true neutrino energy and time-of-arrival at the detectors, selecting distinct energy spectra from a wide-band neutrino beam. This technique complements the “Precision Reaction-Independent Spectrum Measurement (PRISM)” \cite{abud2021deep} method, used by DUNE and other experiments, which samples multiple off-axis measurements to reduce systematic uncertainties in interaction modeling. Uniquely, the stroboscopic approach provides access to true energy information at the Far Detector, something no other part of the DUNE experiment can achieve. 
By choosing later-arriving neutrinos, we can select purer low energy flux, hence better measure $\nu_e$ appearance at the second oscillation maximum \cite{dunecollaboration2016longbaseline} in the Far detector. Also it is possible to separate different components of the beam with timing selections.\\
The mechanism of the stroboscopic approach is as follows. Protons impinge on the target, producing hadrons such as pions and kaons, which decay into muons and neutrinos as shown in Figure \ref{fig:deltat_calc}. The difference between the time of arrival of the neutrino at the detector from a sub-relativistic parent hadron of energy E with respect to a high energy hadron traveling with $\sim$ speed of light (c) is given by $\Delta T$ in Equation \ref{eq:deltat_calc}.
\begin{equation} 
\label{eq:deltat_calc}
\Delta T = \Delta T_{had} + \frac{\Delta L_{\nu}}{c} - \frac{(\Delta z_{had} + \Delta z_{\nu})}{c}
\end{equation}
\begin{equation} 
\label{eq:deltat}
\Delta T \approx \Delta T_{had} - \frac{\Delta z_{had}}{c}
\end{equation}
Equation \ref{eq:deltat_calc} can be written as Equation \ref{eq:deltat} for $\Delta L_{\nu} \approx \Delta z_{\nu}$. 
Here $\Delta L_{had}$ refers to the distance traveled by the hadron after its production to where it decays, $\Delta z_{had}$ is the distance traveled by this hadron along the beam axis, and $\Delta T_{had}$ is the time-of-flight of the hadron in the lab frame. 
In this example, the pion decays to a muon and a muon neutrino. The term $\Delta L_{\nu}$ is the distance to the detector. So, $\frac{\Delta L_{\nu}}{c}$ is the neutrino transit time. The $\Delta z_{\nu}$ term is the distance travelled by the neutrino along the beam axis from its production to the detector, and $\Delta T_{\nu}$ is the neutrino time-of-flight. 
For the sub-relativistic hadrons, $\Delta T$ approaches the hadron decay time in the rest frame at the lowest energies. $\Delta T$ becomes zero as the speed of the low energy hadrons reaches the speed of light. So, a non-zero time spread develops until the sub-relativistic hadrons decay into neutrinos, which will then travel at essentially the speed of light c.\\

However, if neutrinos possess a non-zero mass, the transit time $\Delta T_{\nu}$ would depend on the neutrino's energy due to its slightly subluminal velocity. The total energy of a neutrino can be described by the relativistic equation:

\[
E_{\text{total}} = \sqrt{(pc)^2 + (m_{\nu}c^2)^2}
\]

where:
\begin{itemize}
    \item \( E_{\text{total}} \) is the total energy (which is the sum of the rest mass energy and kinetic energy),
    \item \( p \) is the relativistic momentum,
    \item \( c \) is the speed of light,
    \item \( m_{\nu} \) is the rest mass of the neutrino.
\end{itemize}

For a highly relativistic particle such as a neutrino its kinetic energy is much greater than its rest mass energy.
We can also express the kinetic energy as:
\[
E_{\text{kinetic}} = E_{\text{total}} - m_{\nu}c^2
\]
For low kinetic energies, we need the actual neutrino mass \( m_{\nu} \).
The most recent result on the kinematic search for neutrino mass in tritium decay is from KATRIN \cite{aker2022katrin}, an experiment which has so far found no indication of \(m_{\nu_e} \neq 0\) and sets an upper limit:
\[
m_{\nu_e}^{\text{eff}} < 0.8 \, \text{eV} \quad (90\% \, \text{CL})
\]

\begin{figure}[htpb!]
\begin{center}\setlength{\unitlength}{0.5cm}
\includegraphics[width=.45\textwidth]{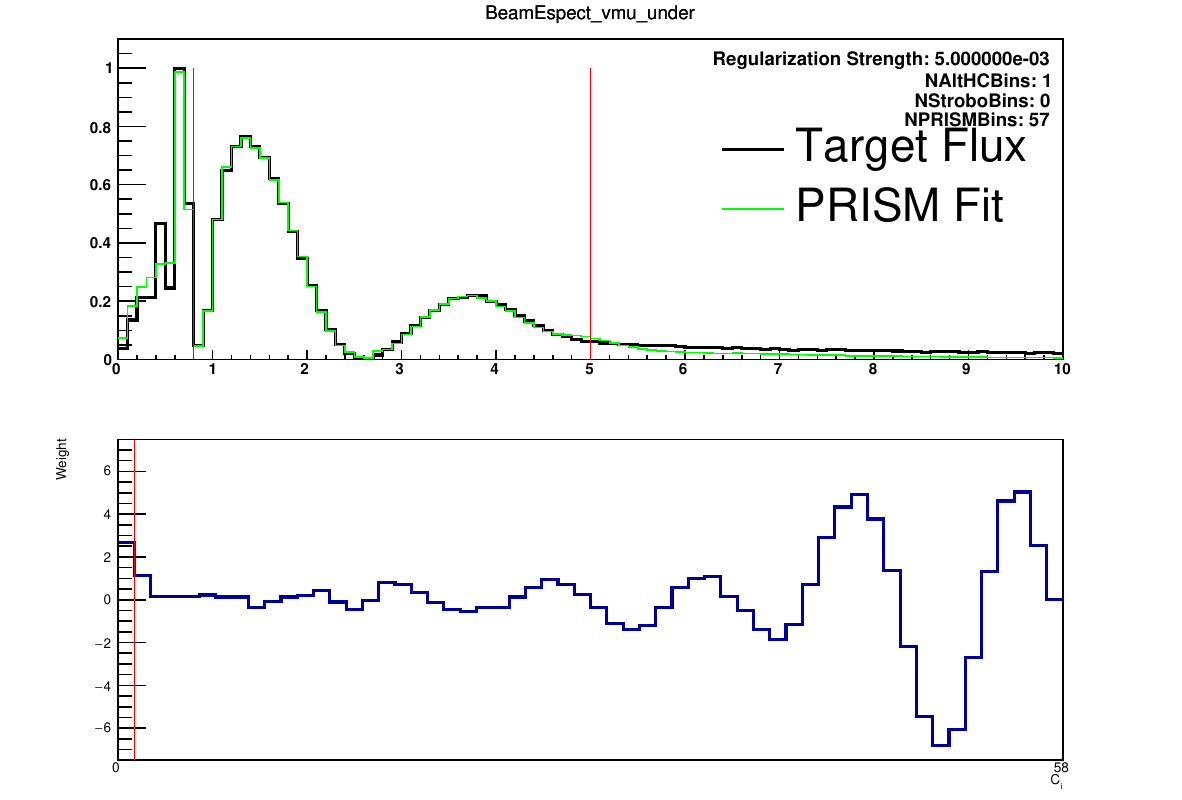}
\includegraphics[width=.45\textwidth]{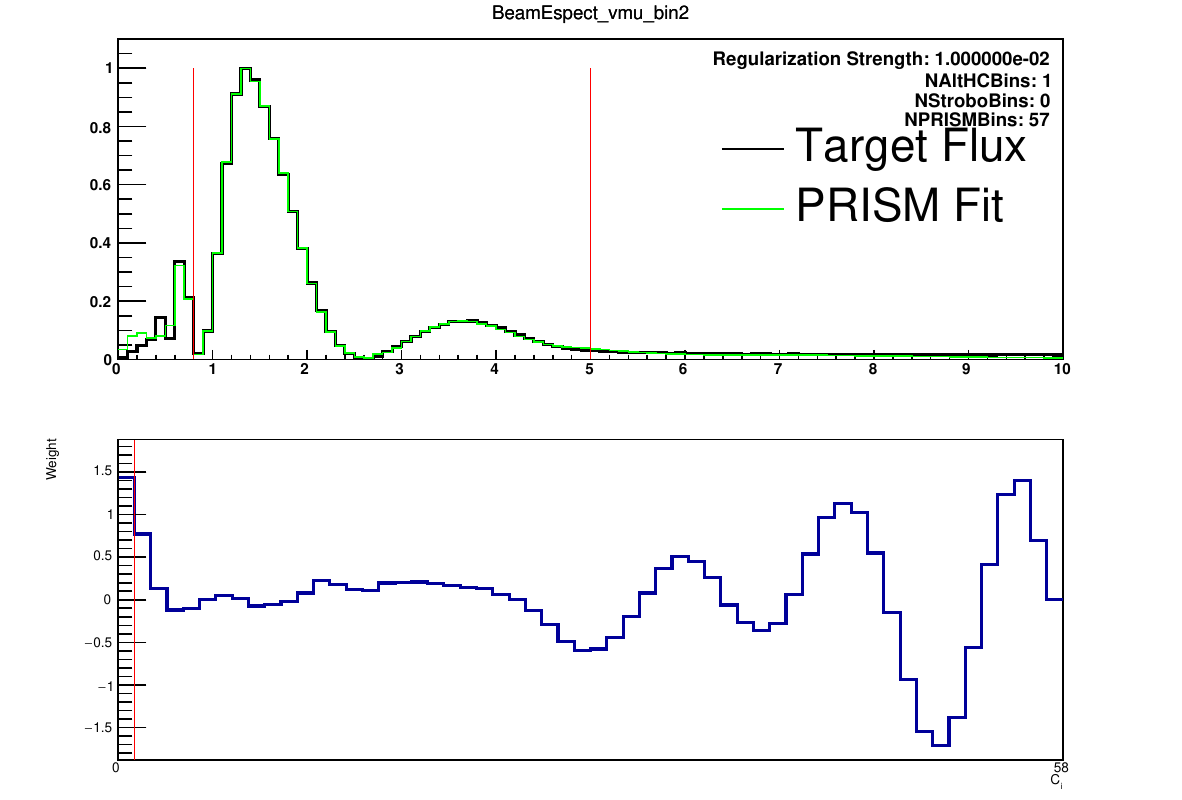}
\caption{PRISM fits to oscillated Far Detector time bins.}\label{fig:PRISMStroboFD}
\end{center}
\end{figure}

\begin{figure}[htpb!]
  \begin{center}\setlength{\unitlength}{0.2cm}
  \begin{tabular}{@{}cc@{}}
   \includegraphics[width=.23\textwidth]{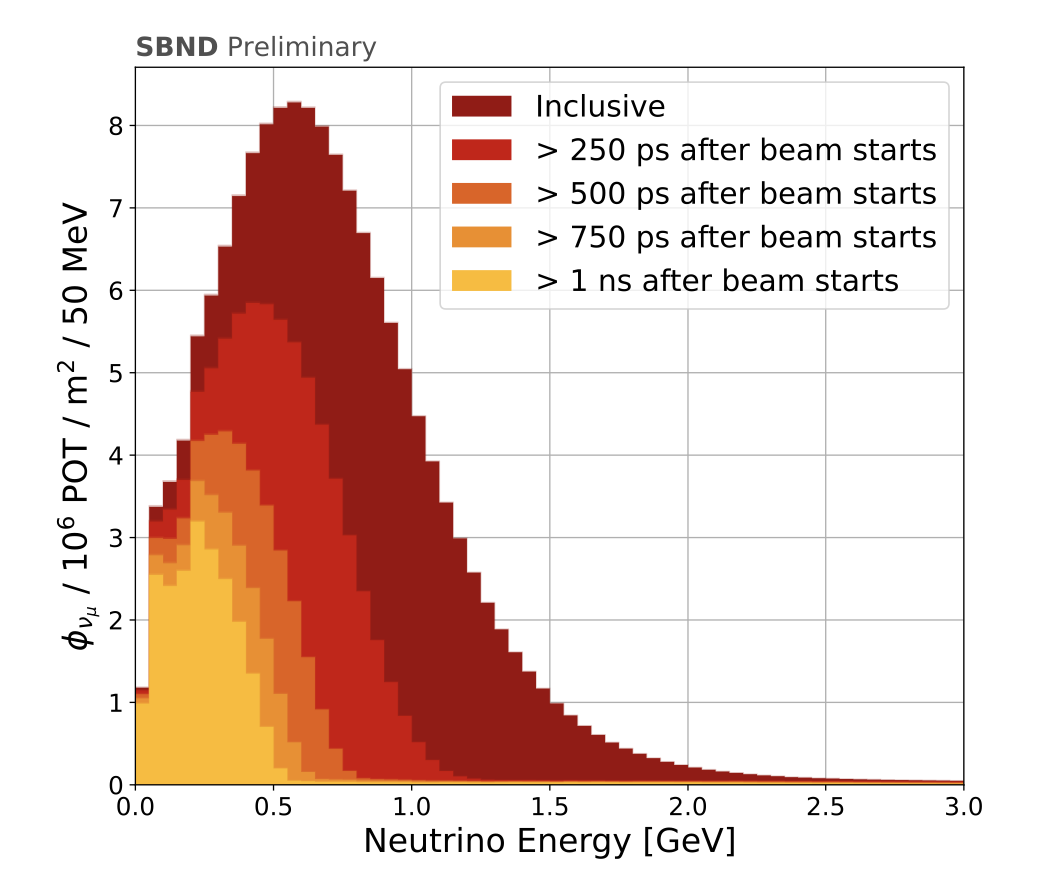} &
    \includegraphics[width=.23\textwidth]{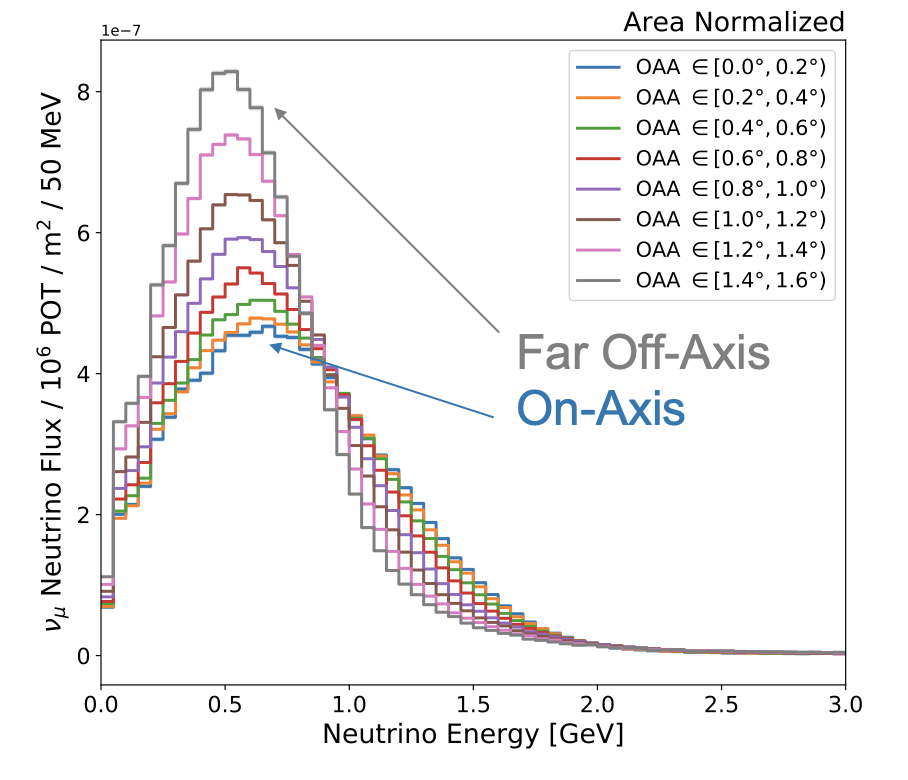} 
    \end{tabular}
\caption{Simulated SBND $\nu_{\mu}$ flux (maroon) overlaid with fluxes corresponding to successively later time selections on the bunch time (left), assuming no time spread of the protons on target: keeping all neutrinos that arrive 250 ps after the start of the neutrino bunch (red), 500 ps after (dark orange), 750 ps (light orange), 1 ns (yellow). SBND-PRISM : $\nu_{\mu}$ flux for different off-axis angles (right). Courtesy: SBND collaboration.}\label{fig:sbndprism}
    \end{center}
  \end{figure}

For the other flavors, the present limits compiled in the listing section of the PDG \cite{pdg2023} read:
\[
m_{\nu_\mu}^{\text{eff}} < 190 \, \text{keV} \quad (90\% \, \text{CL}) \quad \text{from} \quad \pi^{-} \to \mu^{-} + \bar{\nu}_{\mu}
\]
\[
m_{\nu_\tau}^{\text{eff}} < 18.2 \, \text{MeV} \quad (95\% \, \text{CL}) \quad \text{from} \quad \tau^{-} \to n\pi^{+}\nu_{\tau}
\]
\begin{figure}[htpb!]
\begin{center}\setlength{\unitlength}{0.2cm}
\includegraphics[width=.22\textwidth]{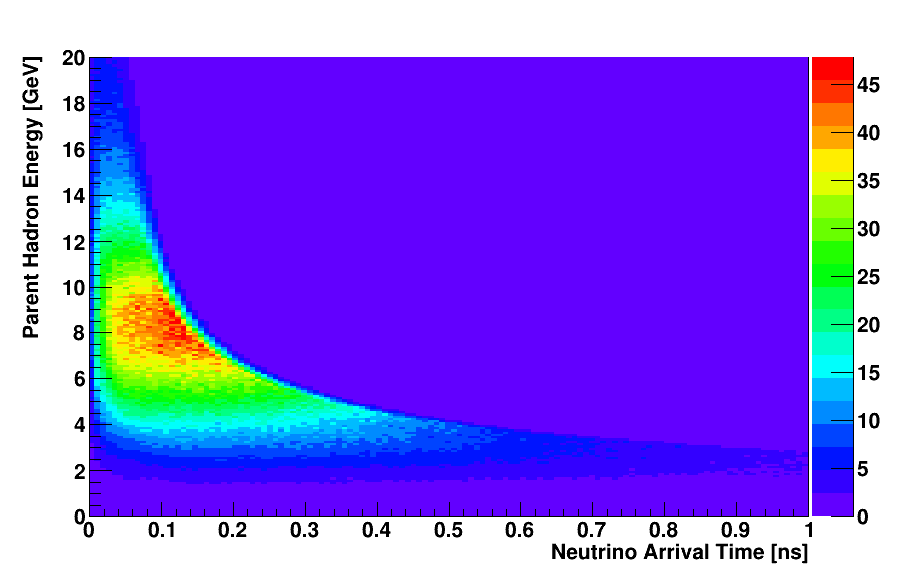}
\includegraphics[width=.22\textwidth]{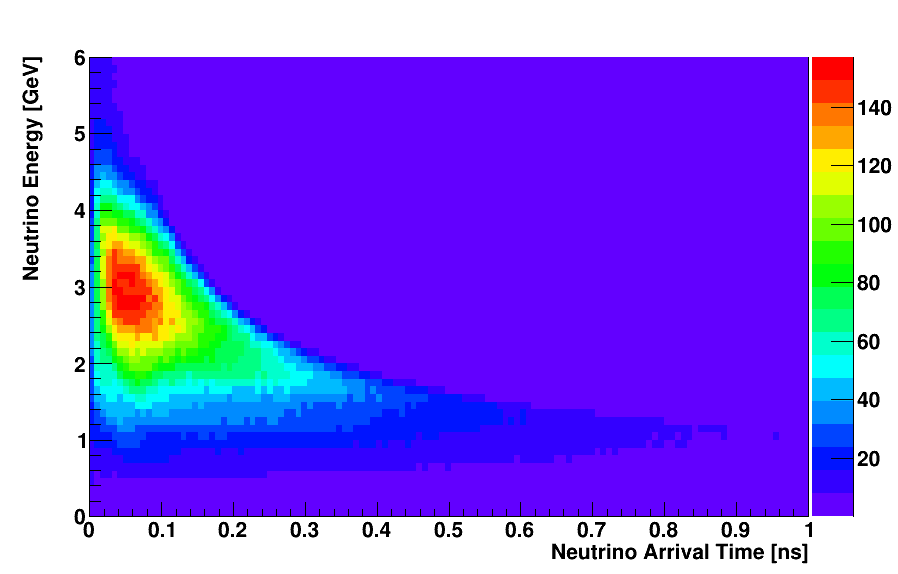}
\caption{\textbf{Left:} distribution of relative neutrino arrival times versus parent hadron energies, \textbf{Right:} similar distribution of relative neutrino arrival times versus neutrino energies for all neutrinos with simulated data of the LBNF beam. All plots are made in Forward Horn Current Mode. The fluxes are plotted in linear arbitary units.}\label{fig:ET}
\end{center}
\end{figure}
The left and the right plots in Figure \ref{fig:ET} show the distributions between $\Delta T$ and all hadron energy and neutrino energy respectively with the simulated data of the LBNF beam \cite{alion2016experiment} where all protons have zero bunch width i.e.\ they strike the target simultaneously. Based on the neutrino arrival time spectrum, the pion energy differentiation is tightly peaked in time and has a wide tail. The systematic uncertainties on the neutrino beam such as the angular and energy uncertainties in the hadron production in the target can potentially affect the correlation between the true neutrino energy and the measured neutrino time-of-arrival and these effects will have to be studied with simulation. However, the magnetic focusing in the beamline will bend the lower energy parent hadrons more than the higher energy ones, thus potentially accentuating the time difference. The longer the hadrons live at high energies, the more restrictive their angular spread is. The beam pipe will also limit the acceptance of the higher energy hadrons. The angular acceptance is even narrower for the Far detector because of its long distance.\\
Timing is also useful for isolating lepton flavor and parent hadron components of neutrino flux from a wide-band beam of neutrinos. Rare $\nu_{\tau}$ candidates are mostly produced from short-lived $\tau$ leptons and their production will be enhanced in the earliest time slice. Hence, application of the stroboscopic approach helps in the $\nu_{\tau}$ energy reconstruction, measurement of $\nu_{\tau}$ charged-current interactions and the associated $\nu_{\mu}$ to $\nu_{\tau}$ oscillation probability at the Far detector.\\
Long baseline neutrino experiments are entering a precision era, and a reduction in the systematic errors to the level of a few percent is necessary to attain their physics goals. LBNF covers a wide range of energies. It enables shape fitting of the oscillation spectrum over a wide range, but the reconstruction of neutrino energy and knowledge of beam composition also present challenges. One of the most challenging sources of these systematic errors arises from the flux along with the neutrino-nucleus interaction cross sections. The dependence of the observable final state particles on the incident neutrino energy is currently not adequately understood. Typically, neutrino energy is determined by observing the final state. However, missing energy from neutrons and nuclear effects can impact the final state. This leads to smearing of $E_{rec}$ relative to $E_{true}$ which is known as the ``feed-down" effect. It is possible for hadrons produced by neutrino interactions to re-interact with the nuclear medium before leaving the nucleus, and the Final State Interactions (FSI) can alter the multiplicity and final state kinematics of the outgoing hadrons. This can lead to the misinterpretation of the primary neutrino interactions. Feed-down effects in the reconstructed neutrino energy are caused by the missing energy from undetected particles such as neutrons and low energy charged pions.\\
Neutrino energy spectra are different at the Near and the Far Detectors. 
Neutrino fluxes are different due to oscillations. The feed-down effects in a Far-to-Near event rate ratio as a function of true neutrino energy do not cancel out. Instead of measuring neutrino fluxes directly, we measure the event rate which is flux multiplied by the interaction cross section. If both the flux and the cross section have uncertainties, we cannot unambiguously tell if we have both correct in our models or both wrong in ways that result in an approximately correct event rate prediction. This becomes exacerbated by reconstructed energy feed-down, the fact that cross-sections are changing rapidly as a function of energy over the DUNE beam energy range, and due to the wide-bandedness of the DUNE beam. Obtaining a better understanding of the cross sections is critical for the DUNE experiment and neutrino physics as a whole. Better knowledge of the neutrino-nucleus cross sections will improve the precision of our oscillation measurements. By identifying reconstruction-independent observables, such as the neutrino arrival time or the off-axis angle positions of the Near Detector, we can isolate different energy spectra within the same beam and effectively disentangle flux and cross-section uncertainties.\\
In the stroboscopic approach, the flux is binned in time, similar to the binning in DUNE PRISM. However, the most unique feature is that timing information can also be applied to the Far Detector. This cannot be done with the PRISM approach.
With timing information, it is possible to divide the oscillated Far Detector spectrum into bins of true neutrino energy narrower than the wide-band unoscillated flux obtained without it. With DUNE PRISM, a mobile Near Detector samples different neutrino energy spectra at different positions and builds up 2D measurements in 2D neutrino flux.\\ 
In order for the PRISM method to work well, one needs to find the fluxes that form a basis that can reproduce any oscillated FD flux. Any discrepancies are filled using Monte Carlo simulations.
\begin{figure}[htpb!]
\begin{center}\setlength{\unitlength}{0.1cm}
\includegraphics[width=0.4\textwidth]{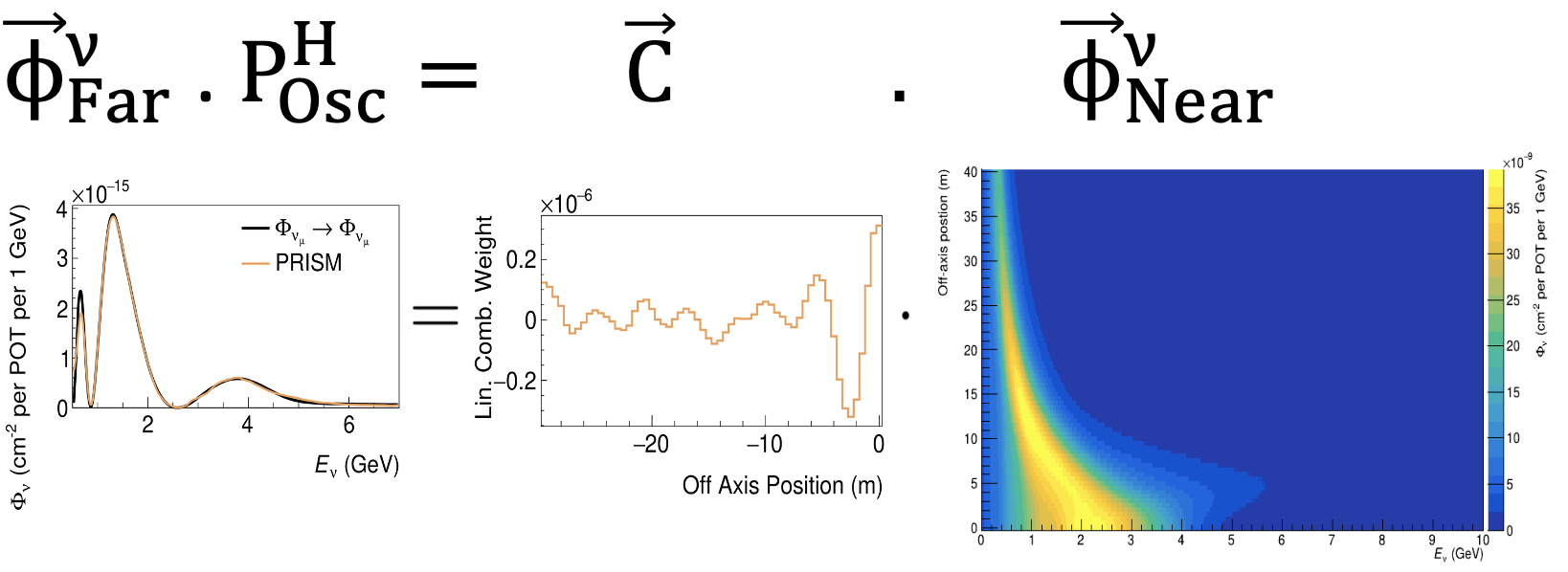}
\caption{Building a Far Detector prediction with PRISM.}\label{fig:PRISM}
\end{center}
\end{figure}
Figure \ref{fig:PRISM} shows that, $\Phi_{Far} (E_{\nu}) . {P_{Osc}}^H (E_{\nu})$ is decomposed into a linear combination of off-axis Near Detector fluxes to measure oscillated flux at the Near Detector. Here $\Phi_{Near} (E_{\nu})$ and $\Phi_\mathrm{Far}(E_{\nu})$ are the fluxes at the Near and the Far Detectors and ${P^H_\mathrm{{OSC}}(E_{\nu})}$ is the oscillation hypothesis.
Weights ($C$) at each off-axis location are solved for. Lastly, the target Far Detector flux is compared with the PRISM fit to the linear combination flux prediction. 
\begin{figure}[htpb!]
  \begin{center}\setlength{\unitlength}{0.5cm}
   \includegraphics[width=0.2\textwidth]{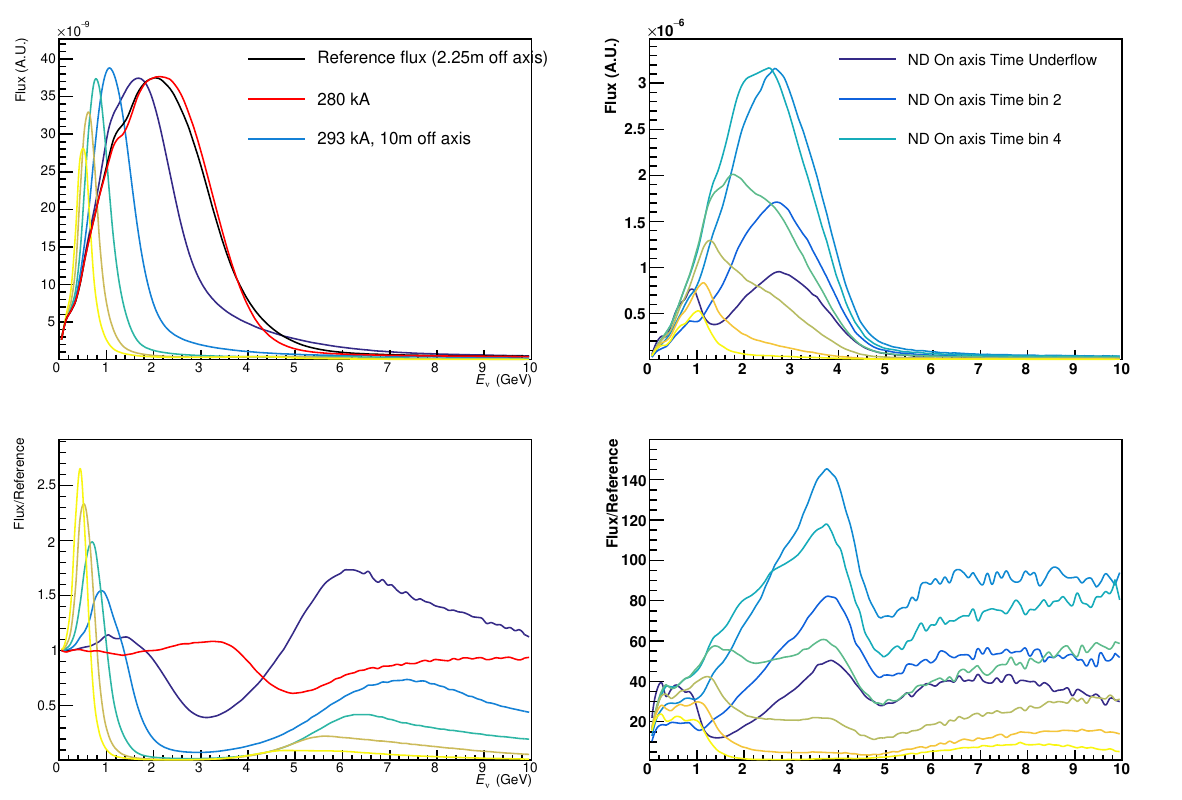}
\caption{Flux components at the Near Detector to perform PRISM fits. \textbf{Top left:} PRISM off-axis and alternate horn current (280 kA) fluxes, \textbf{Top right:} Stroboscopic fluxes, \textbf{Bottom left:} Ratio of the PRISM fluxes and the 280 kA flux to the reference flux, \textbf{Bottom right:} Ratio of the stroboscopic fluxes to the reference flux.}\label{fig:prismfluxes}
    \end{center}
  \end{figure}
PRISM only changes the shape below the flux peak when moving further off-axis. Consequently, fitting the rising edge of the first oscillation maximum (falling edge of oscillated flux at about 3.5 GeV) is quite challenging. 
\begin{figure}[htpb!]
\begin{center}\setlength{\unitlength}{0.5cm}
\includegraphics[width=.45\textwidth]{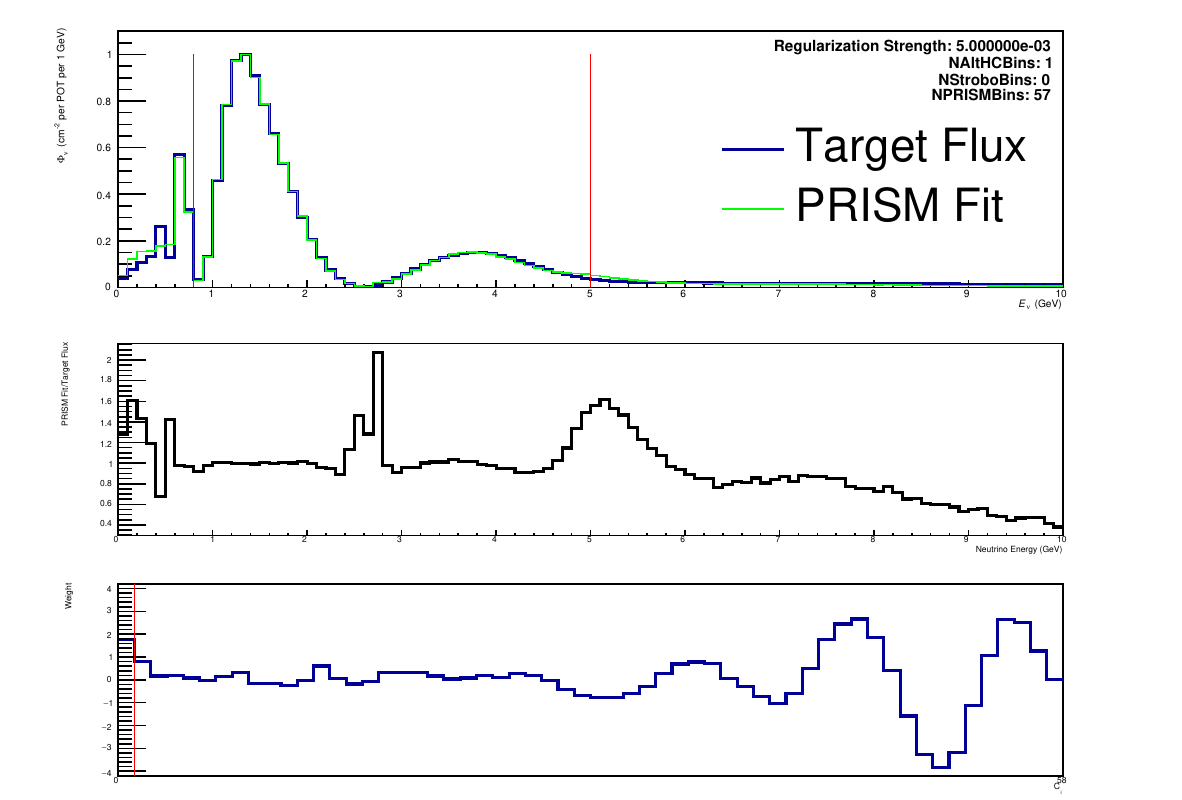} 
\includegraphics[width=.45\textwidth]{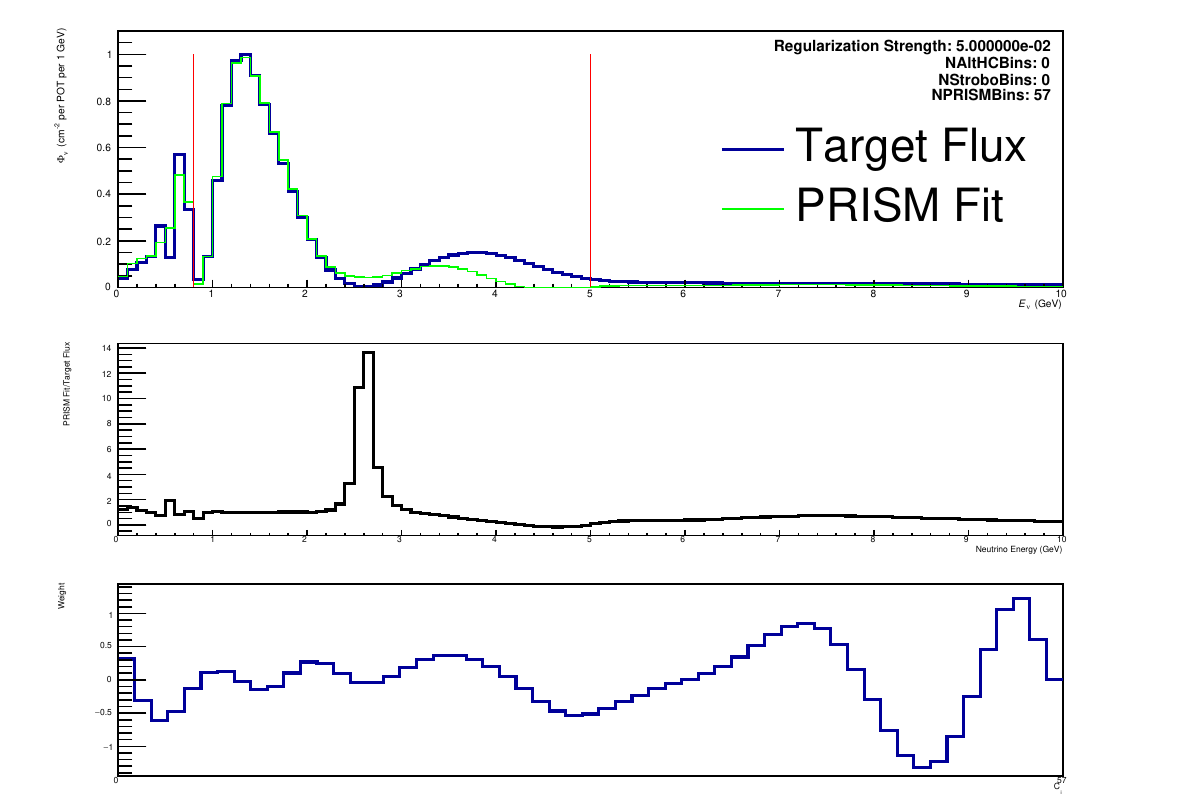}
\includegraphics[width=.45\textwidth]{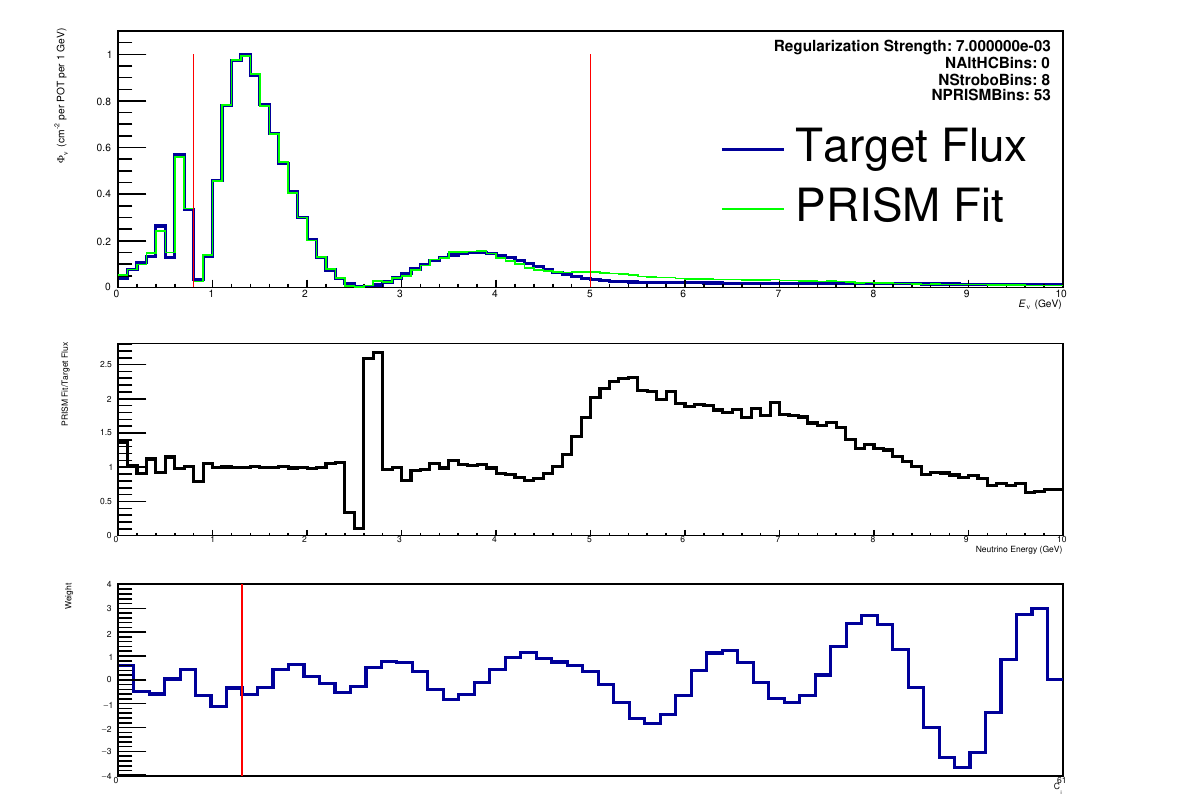}
\caption{PRISM fits. \textbf{Top:} the top panel shows the default PRISM fit with PRISM on and off-axis fluxes and altHC flux. The red lines show the fit region. The middle panel shows the ratio of the PRISM fit to the target flux. The bottom most panel enumerates the coefficients of the nominal on and off axis fluxes to the right of the red line and the altHC coefficients to the left of the red line. \textbf{Middle:} the top panel shows the PRISM fit with PRISM on and off-axis fluxes only, no altHC flux. Same explanations for the middle and the bottom panels. \textbf{Bottom:} the top panel shows the PRISM fit with PRISM off-axis fluxes only, stroboscopic on-axis fluxes and no altHC flux. Same explanations for the middle and the bottom panels.}\label{fig:PRISMStrobo}
\end{center}
\end{figure}
Solution to that is to take a special beam run at a lower alternate horn current. Running at a lower alternate horn current reduces flux at higher energies, but leaves flux below the peak largely the same. Furthermore, including this mode in the PRISM fit provides the control needed to fit the whole spectrum well.\\
As a technique complementary to PRISM, the stroboscopic approach has two obvious advantages. In the first place, the oscillated time integrated spectrum of the Far Detector can be fitted to PRISM and Stroboscopic approaches in the Near Detector and it shows that the PRISM Near Detector program can be further enhanced with a fast Near Detector. By adding the time-sliced Near Detector fluxes along with PRISM fluxes, it is also possible to run without the alternate horn current mode (altHC) that affects the Far Detector flux. Furthermore, this saves valuable operational time.
Figure \ref{fig:prismfluxes} shows the flux components that are used to build the oscillated flux at the Near Detector. In the top left are PRISM off-axis fluxes and the altHC flux (at 280 kA), while the top right shows stroboscopic fluxes in the Near Detector with simulated data for the Forward Horn Current (FHC) mode. The reference flux shown in red is the flux at the 2.25 m off-axis position. In the bottom two plots, we can see the ratio to the reference flux. The shape difference in the bottom right plot (for stroboscopic slices) is promising since it shows differences above the flux peak.\\
A PRISM fit has been performed using the flux components as shown in Figure \ref{fig:prismfluxes}, under three different scenarios: 1) with PRISM and altHC fluxes only, 2) with PRISM fluxes alone without the altHC flux component, 3) replacing the PRISM on-axis contribution with eight time sliced stroboscopic fluxes, while keeping the PRISM off-axis components intact, without the altHC flux component.
Figure \ref{fig:PRISMStrobo} shows in the top plot a default PRISM fit with PRISM on and off-axis fluxes and the altHC flux. This fit is shown in the top panel of the top plot. The red lines show the fit region. The middle planel in the top plot shows the ratio of the PRISM fit to the target flux. The bottom most panel in the top plot shows the coefficients of the nominal on and off axis fluxes to the right of the red line and the altHC coefficients to the left of the red line. 
In the middle plot, the top panel shows the PRISM fit with PRISM on and off axis fluxes only but not altHC flux. In the bottom plot, we add the stroboscopic on-axis time-sliced flux slices onto the PRISM off axis fluxes only but no altHC flux. In the middle fit, without the altHC flux, the fit goes bad, but adding the stroboscopic slices in the bottom plot without the altHC flux makes the fit good again. For the middle and the bottom plots, the same explanation stand for the middle and the bottom panel in each case.
Figure \ref{fig:PRISMStrobo} illustrates three PRISM fits. In the top plot, the default PRISM fit uses both on-axis and off-axis fluxes, along with altHC flux. The red line indicates the fit region. The middle panel shows the ratio of the PRISM fit to the target flux, while the bottom panel displays the fit coefficients, with nominal on-axis and off-axis fluxes on the right of the red line and altHC flux coefficients on the left.\\
In the middle plot, the fit only uses PRISM on-axis and off-axis fluxes, excluding the altHC flux. Without the altHC flux, the fit quality deteriorates.\\
The bottom plot shows the fit using stroboscopic on-axis time-sliced fluxes with off-axis fluxes, but no altHC flux. Adding the stroboscopic slices improves the fit, restoring its quality. The middle and bottom panels in both the middle and bottom plots serve the same purpose as in the top plot. The possibility of performing PRISM fits without the altHC shows promise.\\
The stroboscopic approach also has PRISM complementarity in that it can provide Far Detector oscillated time slices, which means it should operate with the default PRISM program if a fast Far Detector is available.\\
Instead of using the full oscillated, time-integrated Far Detector flux, multiple PRISM fits have been performed in this study to fit oscillated FD time bins that are obtained from stroboscopic slices as shown in Figure \ref{fig:PRISMStroboFD}.
Two time bins are shown in this figure. This shows that we can extend PRISM analysis if we have a fast Far Detector. DUNE ND is expected to have neutral lepton sensitivities that are competitive with other experiments and will be enhanced by the application of stroboscopic techniques. By improving timing at the DUNE Far Detector, we can distinguish inelastic interactions of cosmogenic inelastic Boosted Dark Matter (iBDM), which would produce a prompt proton track followed by an electron$/$positron pair with a displaced vertex, from cosmic ray and atmospheric neutrino backgrounds, while improving detection efficiency for the signal. In addition, fast timing could be used in large liquid argon detectors to detect nuclear de-excitations of 40Ar following a baryon number violating process in order to increase efficiency and reduce background in search for proton decay \cite{DUNEBSM}.\\
LBNF Near and Far detectors can be used to divide neutrino flux by neutrino arrival time by creating short proton bunches in MI if the detectors have a time resolution similar to the proton bunches at the target. Though time resolutions of (O(100 , \text{ps})) are attainable in the water Cherenkov detectors, liquid Argon TPCs as currently conceived are slow detectors. At present, the MicroBooNE detector intrinsic resolution is 2.2 ns \cite{muboone}. A few ns of resolution is expected via the photon detectors in DUNE Liquid Argon detectors, but this could be addressed with minor upgrades. Getting the interaction time precisely requires a precise measurement of prompt light at one point along the track, provided the rest of the track is constructed properly.\\
The optical properties of liquid Argon detectors are similar to those of water. Combined with electron drift, Cherenkov light provides an excellent handle for establishing T0. From the electron TPC data, liquid Argon-based detectors are able to precisely reconstruct each event in space. It is possible to use the reconstructed track to simulate the detected time and position of Cherenkov photons emitted by some or all of the charged particles. Only one parameter namely the neutrino event time needs to be fitted for, in the comparison of the 4D-coordinates of simulated photons and measured photons \cite{angelico2020measuring}. A silicon photomultiplier or a large area picosecond photodetector (LAPPD) can provide the necessary time and space resolution.\\
The application of the stroboscopic approach is also of interest to the Short-Baseline Near Detector (SBND) which will be located 110 m from the Booster Neutrino Beam (BNB) neutrino source. The transverse size of SBND detector limits the range of off-axis angles. Unlike DUNE, they will not move the detector. Figure \ref{fig:sbndprism} \cite{sbndprism} demonstrates how this approach could optimize the selection of the neutrino flux in SBND based on time slicing and off-axis angles. The left plot shows the time-sliced SBND fluxes peaking at lower energies with later time-cuts on neutrino arrival time whereas the right plot shows flux spectra chosen by different off-axis angles. By selecting different beam fluxes, the stroboscopic approach could enhance the search for new physics in SBND.

\section{Importance of Bunch Length}
\begin{figure}[htbp]
    \centering
    \includegraphics[width=0.45\textwidth]{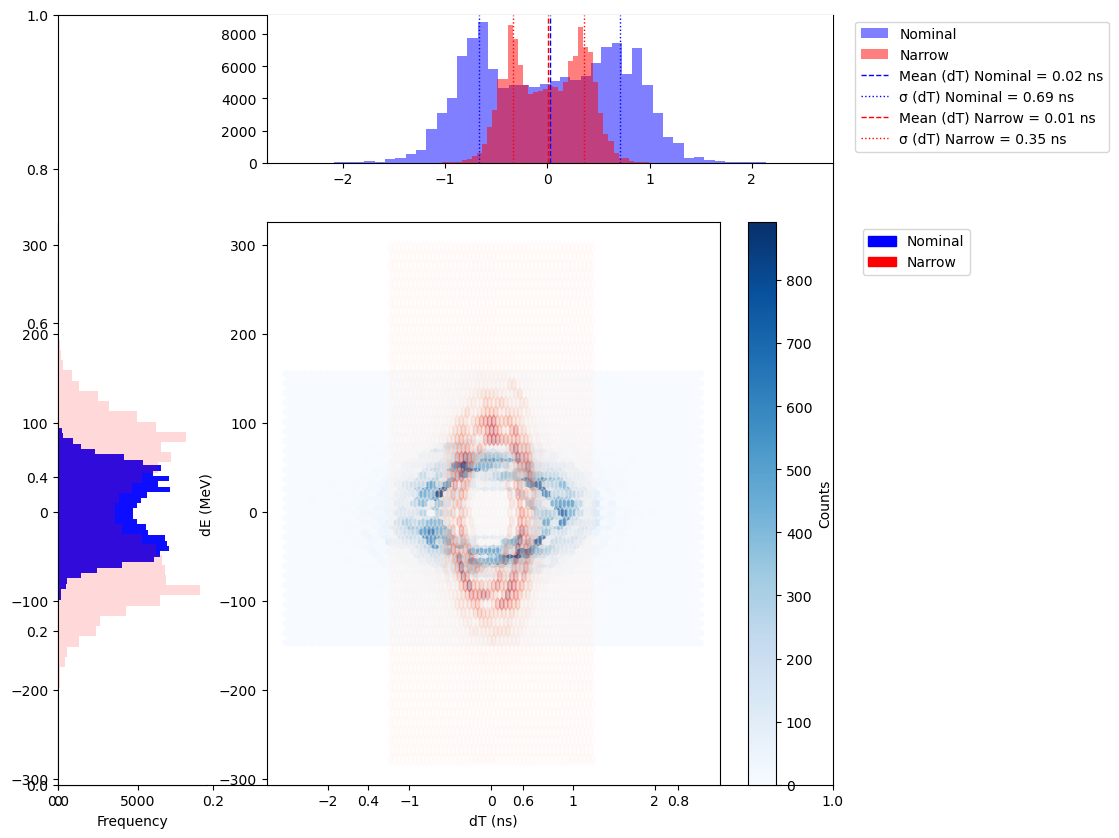}
    \caption{Simulation of phase space with nominal (blue) and minimized (red) bunch lengths. The plot shows the exchange of time (dT) and energy (dE) with varying bunch lengths in the Main Injector at Fermilab. The conservation of phase space \(\Delta t \cdot \Delta E\) allows for flexibility in managing the bunch characteristics.}
    \label{fig:phasespace}
\end{figure}

\begin{figure}[htpb!]
  \begin{center}\setlength{\unitlength}{0.2cm}
  \begin{tabular}{@{}cc@{}}
   \includegraphics[width=.23\textwidth, clip, trim=5 5 5 70]{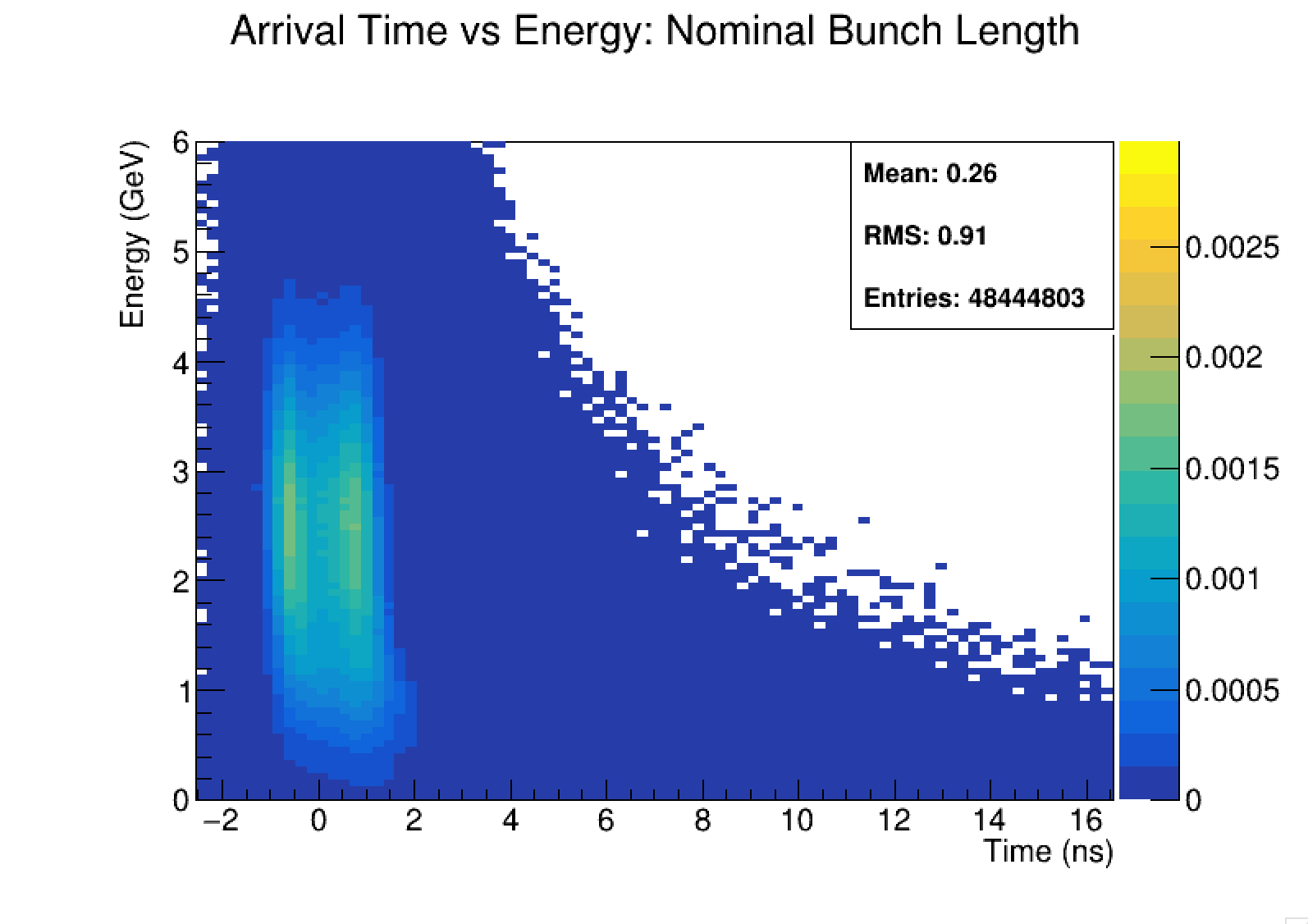} &
    \includegraphics[width=.23\textwidth, clip, trim=5 5 5 70]{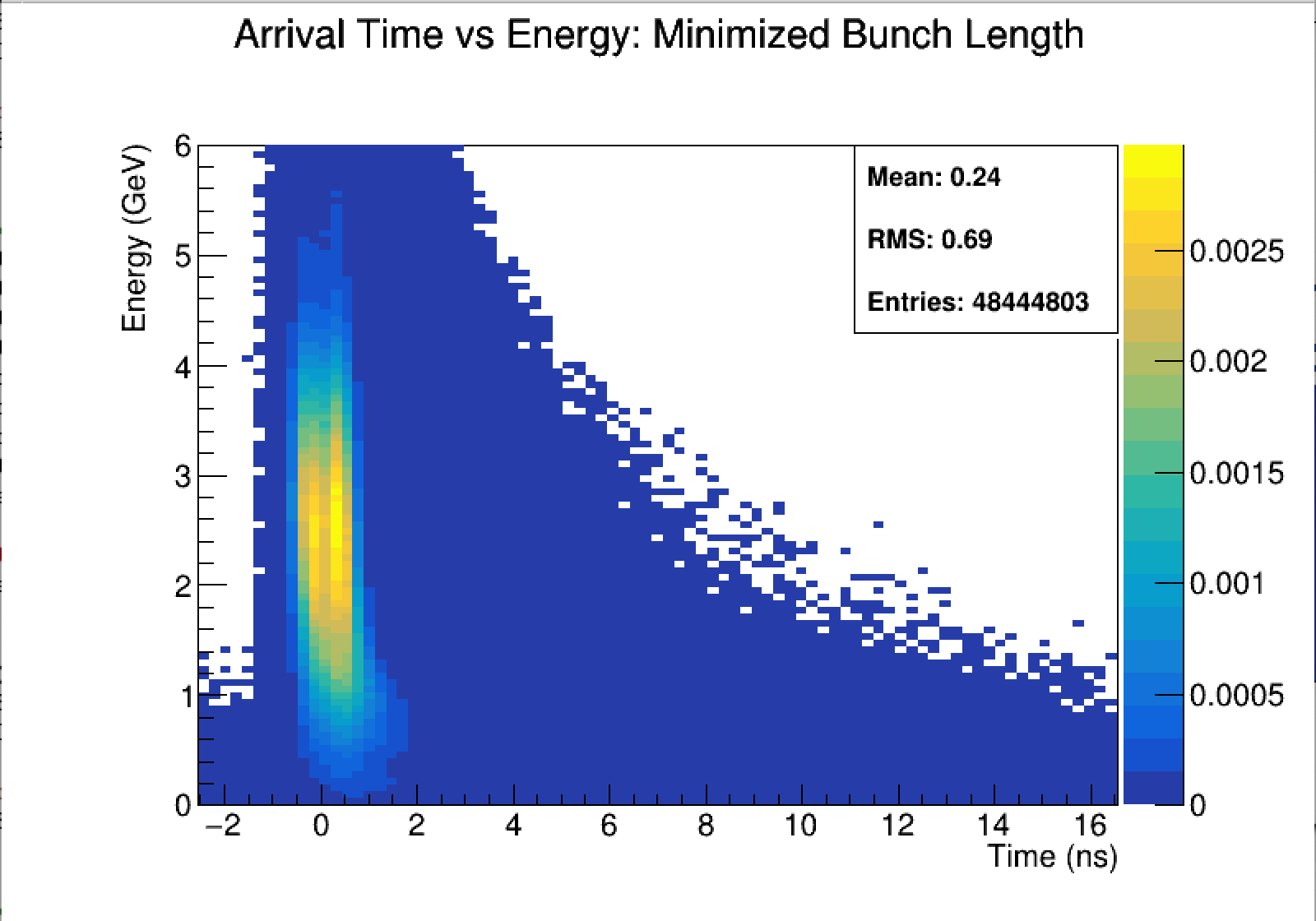} 
    \end{tabular}
    \caption{Arrival Time vs. Energy plots comparing nominal (left) and minimized (right) bunch lengths. These results are based on simulations from DUNE and the BLonD code, which models the Main Injector proton beam time structure. The minimized bunch length improves time correlation, enhancing the precision of the stroboscopic approach.}
    \label{fig:tvse}
    \end{center}
\end{figure}

The length of proton bunches is a key factor in neutrino production and detection. While a proton bunch with zero width is unrealistic, having a finite bunch width means the protons responsible for producing neutrinos are spread over a range of times. This spread causes the production times of the neutrinos to smear out, and the time resolution of the detector further degrades this precision.\\
Figure \ref{fig:tvse} shows the relationship between neutrino arrival time and energy for two cases: nominal and minimized bunch lengths in the Main Injector at Fermilab. In the nominal mode, with a RMS bunch length of $\Delta t = 0.69$ ns, time correlations become less distinct, and bunches can overlap, complicating accurate detection. By shortening the RMS bunch length to $\Delta t = 0.35$ ns using bunch rotation, the timing becomes more precise, enhancing the precision of the stroboscopic approach.\\
As shown in Figure \ref{fig:phasespace}, the phase space, represented as $\Delta t \cdot \Delta E$, remains conserved throughout this process, enabling an efficient exchange between time and energy. Importantly, this exchange can occur in either direction, providing flexibility in controlling the bunch characteristics.\\
The phase space shown in Figure \ref{fig:phasespace} 
is a result of the slip-stacking process \cite{eldred2014slipstacking}, where two beams with slightly different momenta are combined in the Recycler using two RF systems running at different frequencies. Due to the small energy difference, the bunches ``slip" past each other in phase space over time. Eventually, the bunches are merged to form a higher intensity beam. In slip-stacking, the manipulation of two distinct momentum populations creates a complex, non-Gaussian phase space structure.\\

The phase space becomes stretched and rotated, as the different momentum components occupy different regions in energy-time or energy-position space. Once the two bunches are merged in the Recycler and injected into the Main Injector for further acceleration, the resulting distribution in phase space takes on a ``donut-shaped" pattern.\\
The longitudinal dynamics code BLonD \cite{osti_1358085} models the proton bunch time structure in the Main Injector. The flux has been simulated for the Forward Horn Current (FHC) operation of the LBNF beam.
The simulations show the expected neutrino flux at the near detector, with bunch rotation performed in the Main Injector for the minimized bunch length scenario.\\

\begin{figure}[htpb!]
  \begin{center}\setlength{\unitlength}{0.2cm}
   \includegraphics[width=.33\textwidth, clip, trim=5 5 5 70]{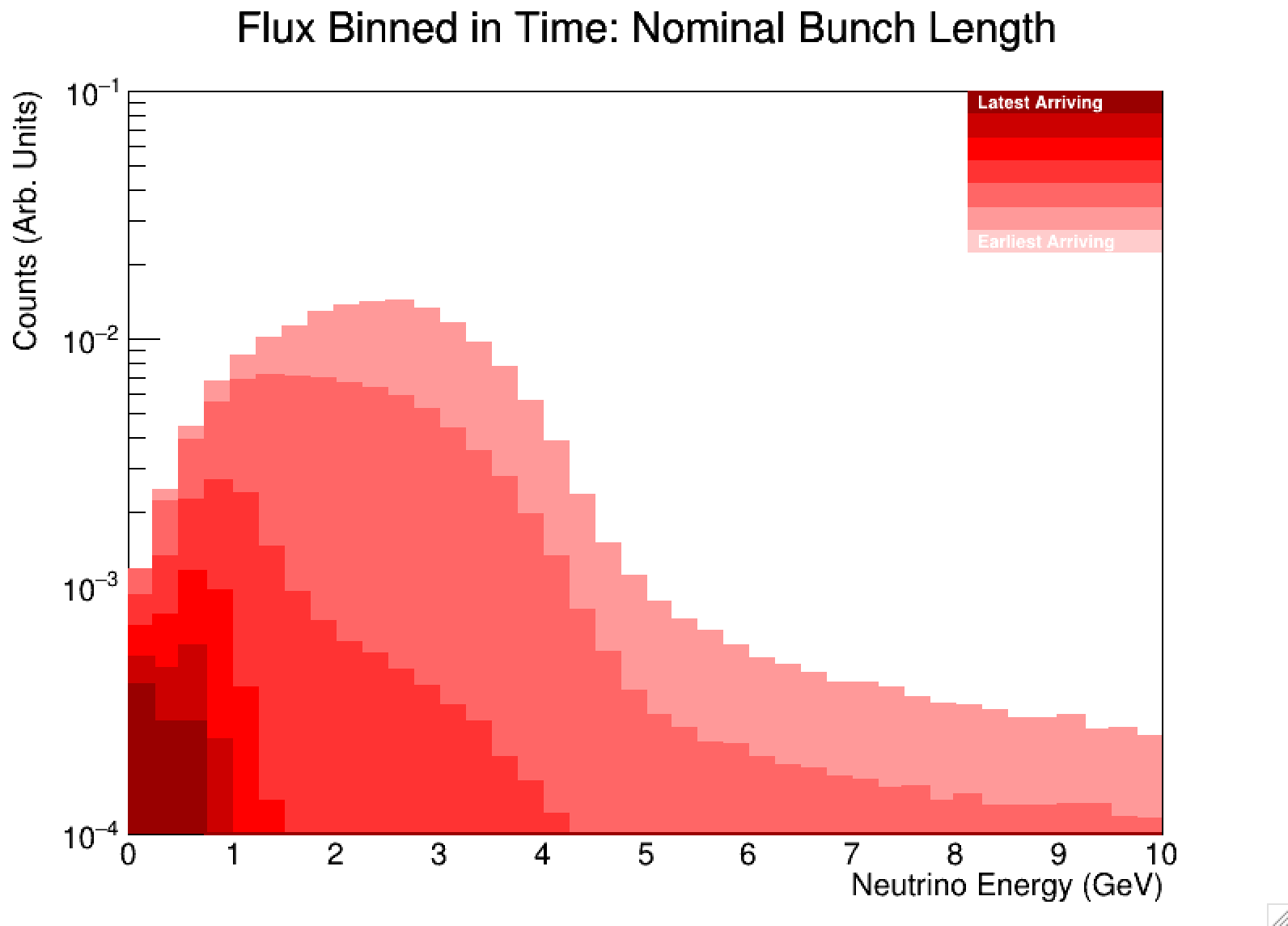}
    \caption{The simulated DUNE forward horn current  fluxes corresponding to increasingly later time-cuts on the bunch time with the minimized bunch length of $\Delta t = 0.35$ ns.}
    \label{fig:fluxbins}
    \end{center}
\end{figure}

\begin{figure}[htbp]
    \centering
    \includegraphics[width=0.45\textwidth]{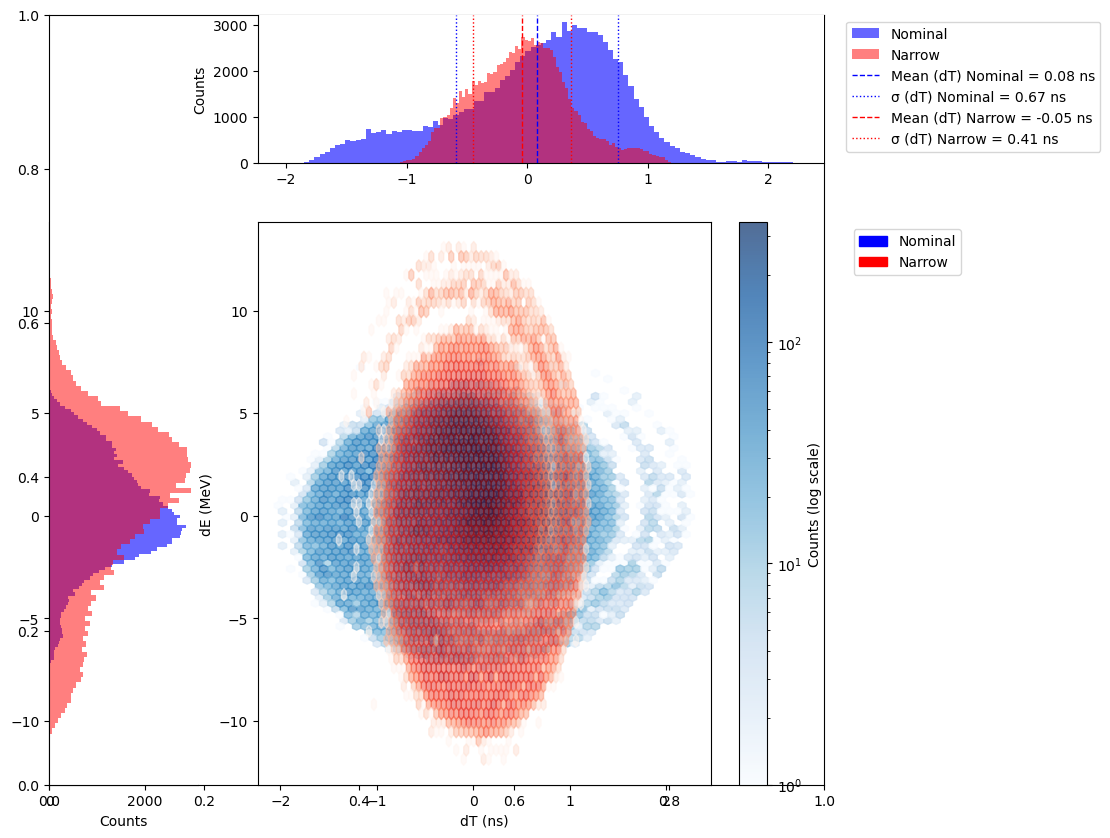}
    \caption{Simulation of phase space with nominal (blue) and minimized (red) bunch lengths. The plot shows the exchange of time (dT) and energy (dE) with varying bunch lengths in the Booster at Fermilab.}
    \label{fig:phasespace_booster}
\end{figure}

Time-slicing based on the arrival of the proton bunch generates distinct neutrino energy spectra for each time slice. Each time slice corresponds to a bin of 250 ps, starting from 500 ps prior to  \( t=0 \). This method is illustrated in Figure \ref{fig:fluxbins}, which presents the simulated beam energy spectrum for different time selection cuts relative to the start of the neutrino bunch. 
Currently, the operational proton beam at the Main Injector typically has an RMS bunch length of around $\sim$1 ns.\\
One approach to reduce the bunch length is to superimpose a higher-frequency RF structure (e.g., the 10th harmonic of 53.1 MHz) onto the proton bunch after acceleration but before extraction. This could potentially bring the RMS bunch length down to around 100 ps, as suggested in \cite{PhysRevD.100.032008}. Achieving this would involve adding a high-frequency cavity, such as a 500 MHz Cornell B-cell cavity \cite{Chao:1490001}, though this would require significant investment. Alternatively, we propose using bunch rotation in both the Main Injector and Booster to achieve a comparable sub-ns bunch length, offering a more cost-effective solution.\\
Figure \ref{fig:phasespace_booster} shows the simulated longitudinal phase space of the Booster beam in both nominal and narrow bunch modes, achieved through bunch rotation. In the nominal mode (blue) the beam shape that is optimized for efficient injection into the Recycler for slip stacking. To achieve this, the momentum spread of the beam is reduced by manipulating its longitudinal phase space distribution. 

In the narrow bunch mode (red), the phase space distribution is rotated, and the beam is extracted when its longitudinal width is minimized. The bunch rotation process results in non-Gaussian energy and time profiles. In the narrow bunch mode, the RMS bunch length from 0.67 ns to 0.41 ns, while elongating the energy distribution. This process creates a more compact beam in time, improving timing precision for downstream processes such as neutrino detection. These results, based on BLonD simulations, demonstrate that bunch rotation provides a cost-effective method to achieve sub-nanosecond bunch lengths in the Booster, paving the way for the application of the stroboscopic approach in short-baseline neutrino experiments at Fermilab.

While the current bunch lengths in both the Main Injector and Booster are longer than ideal for the stroboscopic approach, ongoing efforts to shorten them are crucial for improving the precision of both long and short baseline neutrino experiments at Fermilab. These advancements will enhance the accuracy and effectiveness of future neutrino detection and measurement techniques.

\section{Bunch Rotation Methodology: RF Manipulation Techniques}
In this study, we applied a bunch manipulation technique known as "bunch rotation" to create a narrow bunch spread in the Booster. The Accelerator Neutrino Neutron Interaction Experiment (ANNIE), a water Cherenkov detector at Fermilab, is already equipped with a fast photodetector system capable of providing the necessary time resolution for this technique. Located downstream of the Booster Neutrino Beam (BNB), ANNIE motivates the use of bunch rotation in the Booster as part of a first proof-of-principle demonstration of the stroboscopic approach. Due to the lower energy of neutrinos in the BNB, the stroboscopic effect is more pronounced and can be observed with bunch lengths on the order of 1 ns, as shown in Figure \ref{fig:sbndprism}, although shorter bunch lengths are preferable.\\
Currently, the Booster beam is bunch rotated before being sent to the BNB line and Recycler ring, but this process is typically used to elongate the bunch spread. In contrast, we performed bunch rotation in the Booster to achieve a shorter bunch spread. Additionally, the use of narrow proton bunches could benefit a fast Near Detector capable of scanning the off-axis neutrino beam, enhancing its sensitivity to light dark matter searches \cite{De_Romeri_2019}. MiniBooNE has recently utilized timing in several analyses to select stopped kaons from the NuMI beam and to search for heavy dark matter particles \cite{PhysRevD.98.112004}.\\
In a recent beam study, our goal was to achieve the narrowest possible proton bunch lengths within the Fermilab Booster by adjusting key parameters such as extraction time, beam intensity, and RF frequency. These efforts focused on optimizing the bunch rotation process, which is essential for implementing the stroboscopic approach in neutrino experiments.

\begin{figure}[htpb!]
  \begin{center}\setlength{\unitlength}{0.2cm}
    \includegraphics[width=.30\textwidth]{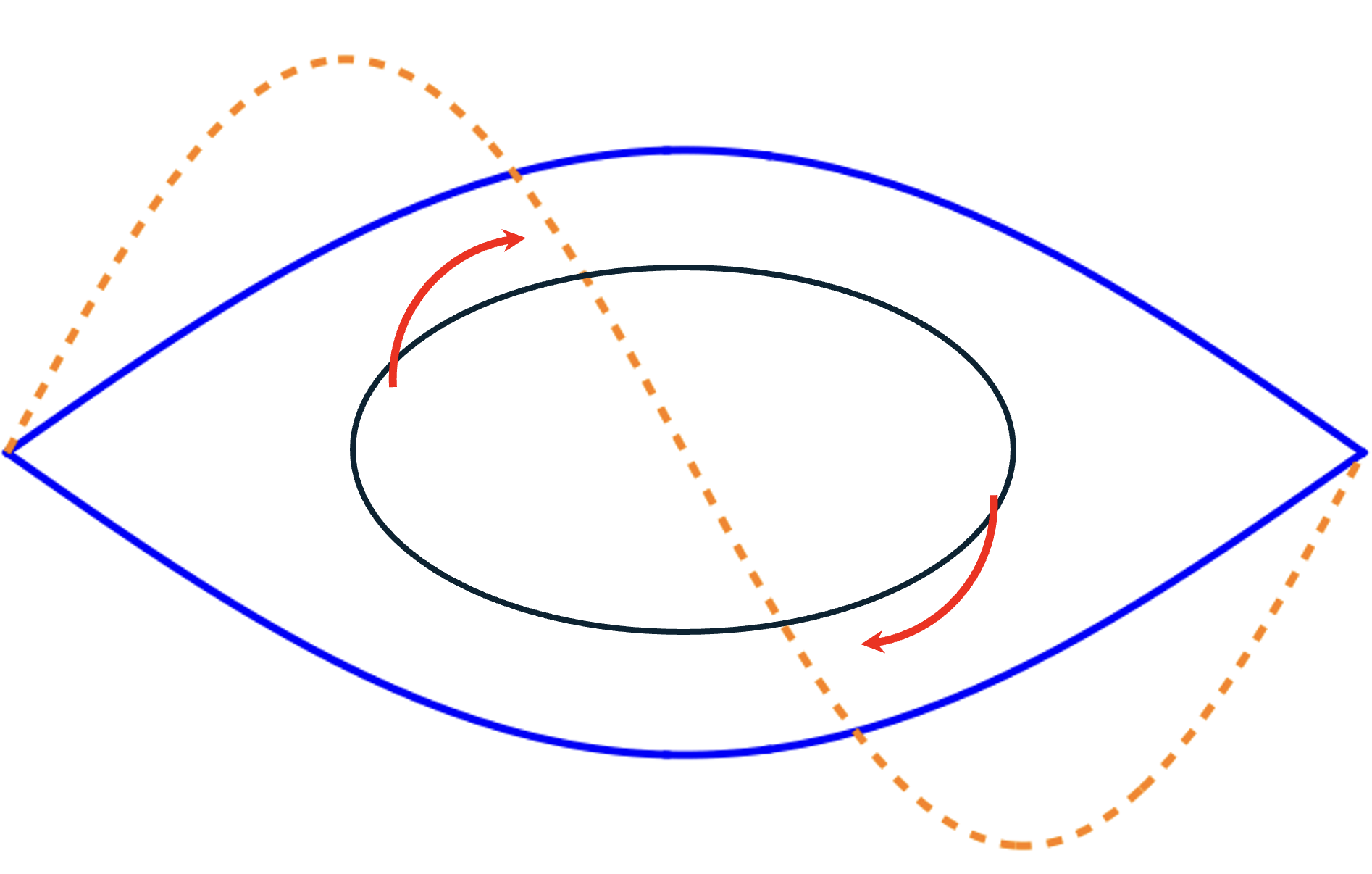} 
    \caption{Concept of longitudinal phase rotation using QBR.}
    \label{fig:qbr}
    \end{center}
\end{figure}

The Fermilab Booster is a rapid-cycling synchrotron ring with a 474.2-meter circumference, equipped with 22 RF cavities and 96 combined-function magnets. Each cavity can deliver up to 50 kV of RF voltage, with the RF frequency modulated from 38 to 53 MHz as the beam energy increases from 0.4 to 8.0 GeV. This modulation is critical to managing the beam’s longitudinal dynamics as it accelerates, ensuring the beam can be properly shaped for experiments downstream.
\begin{figure}[htpb!]
  \begin{center}\setlength{\unitlength}{0.2cm}
    \includegraphics[width=.51\textwidth]{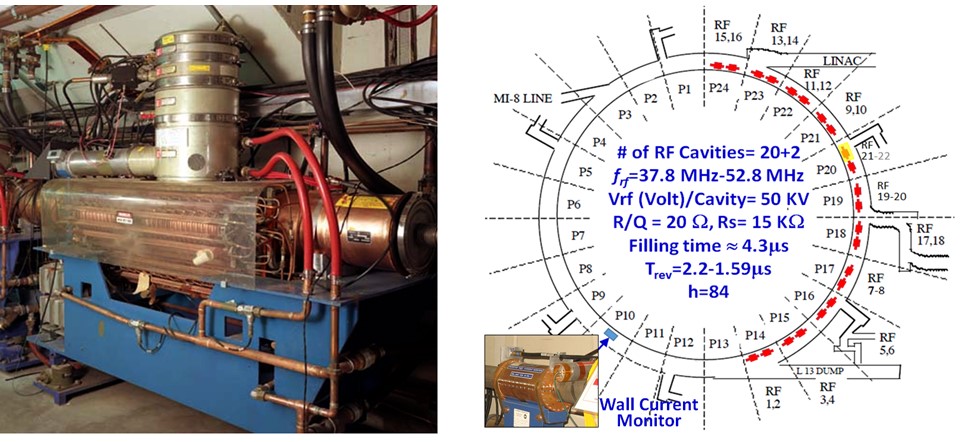} 
    \caption{An example of a refurbished Booster RF cavity (left) and schematic view of the Fermilab Booster ring indicating the distributions of RF cavities (right). Location of the wall current wall monitor used for measuring the line- charge distributions of the beam pulses in the Booster is also shown~\cite{Bhat2017}.}
    \label{fig:booster_layout}
    \end{center}
\end{figure}
Figure~\ref{fig:booster_layout} shows the Booster RF cavity and the layout of the Booster RF cavity and the resistive wall current monitor (RWM) in the Booster Ring. 
To rotate the Booster beam bunch, an RF electric field at twice the synchrotron frequency is applied in the Booster RF cavities, superimposed on the nominal stationary RF bucket. This technique, known as Quadratic Booster Rotation (QBR), induces a longitudinal phase rotation that affects both the bunch length and the momentum spread of the Booster beam at the doubled synchrotron frequency \cite{Eldred2021} (see Figure~\ref{fig:qbr}. 
This figure illustrates this process, with the longitudinal bunch length (in ns) on the x-axis and the momentum spread (in MeV/c) on the y-axis. The blue solid curve outlines the RF bucket boundary, representing the stable region for synchrotron oscillations. The black solid curve shows the trajectory of a beam particle undergoing small-angle synchrotron oscillations within the bucket, while the dashed red line represents the RF waveform applied during QBR.
Consequently, the final bunch length in the Booster ring is determined by the timing of the Booster beam extraction. Under nominal beam operation, the QBR technique is used to elongate the bunch length to facilitate slip stacking in the Recycler Ring for 120 GeV neutrino oscillation experiments. However, this study demonstrates the ability to achieve shorter bunch lengths using QBR by rotating the beam's phase space.

As shown in the simulated phase space plots (Figure~\ref{fig:phasespace_booster}), by adjusting the beam extraction timing and advancing the synchrotron phase by $\pi/2$ or $3\pi/2$ from the nominal operation, we effectively rotate the phase space of the Booster beam. This phase space rotation compresses the longitudinal bunch length while increasing the momentum spread. In nominal operation, the bunch is elongated for slip stacking, resulting in a broader distribution in the phase space. However, when bunch rotation is applied, the phase space is manipulated to achieve a shorter bunch length, as seen in the narrow distribution of the simulation results.
The shortened bunches are then extracted from the ring and delivered to the BNB target through the 8-GeV beam line and the BNB beam line. 

\section{Bunch Rotation Methodology: Data Acquisition Process}
The proton beam was directed through the Booster Neutrino Beam (BNB) line to detectors SBND and ANNIE. This work was conducted entirely within the Booster and not the Main Injector.
The process began with the nominal Booster beam, where the bunch length was measured at various extraction times under different beam intensities. 
To minimize beam losses that would eliminate high momentum particles, both MI-8 collimators (836 and 838) were initially positioned out of the beamline. As the study progressed, the collimators were gradually moved into the beamline to investigate their effect on the bunch length.
The scans were conducted at two primary intensities: 5E12 (high) and 3E12 (medium) protons per pulse (ppp) as shown in table \ref{tab:bunch_rotation}.\\
\begin{table}[h!]
\centering
\resizebox{0.5\textwidth}{!}{ 
\begin{tabular}{|c|c|c|c|}
\hline
\textbf{Time (AM)} & \textbf{Intensity (E12)} & \textbf{Collimator Position} & \textbf{Bunch $\Delta t$ (ns)} \\ \hline
9:07  & 5.0 & Out & 2.1 (33890 $\mu$s) \\ \hline
9:15  & 5.0 & Out & 1.2 (33884 $\mu$s) \\ \hline
9:21  & 5.0 & Out & 0.9 (33856 $\mu$s) \\ \hline
9:28  & 5.0 & Out & 0.9 \\ \hline
9:35  & 5.0 & Out & 2.1 (33980 $\mu$s) \\ \hline
9:38  & 3.0 & Out & 1.725 \\ \hline
9:41  & 3.0 & Out & 0.75 (33864 $\mu$s) \\ \hline
9:45  & 3.0 & 100 mils in & 0.75 \\ \hline
9:48  & 5.0 & 100 mils in & 0.975 \\ \hline
9:52  & 5.0 & Another 50 mils in & 0.9 \\ \hline
9:57  & 5.0 & Col 836: 50 mils out, 838: in & Beam trip, revert to 9:52 \\ \hline
10:16 & 3.0 & Col 836: 50 mils out, 838: in & 0.825 (33876 $\mu$s) \\ \hline
10:21 & 5.0 & Col 836: 50 mils out, 838: in & 2.1 (33890 $\mu$s) \\ \hline
\end{tabular}
}
\caption{Summary of bunch length measurements at various extraction times and collimator positions.}
\label{tab:bunch_rotation}
\end{table}

\textbf{At high intensity:}\
The nominal bunch length was measured as $\Delta t$ = 2.1 ns with an extraction time of 33890 \(\mu\)s.
Bunch rotation was then applied, and the bunch length was shortened to $\Delta t$ = 0.9 ns at an optimized extraction time of 33856 \(\mu\)s.\\
\textbf{At Medium Intensity:} \\
The nominal bunch length was $\Delta t$ = 1.725 ns with an extraction time of 33890 \(\mu\)s. \\
After bunch rotation, the bunch length was reduced to $\Delta t$ = 0.75 ns at an extraction time of 33864 \(\mu\)s.\\
During these scans, various parameters were monitored, including the intensity monitor signals (BCHG0, BLMS06), bunch-by-bunch beam variations, and signals from both the BNB RWM and pre-target RWM. This data was essential in understanding the beam’s behavior under different conditions and optimizing the bunch rotation process.\\
\textbf{Initial Observations:}
When the bunch length was reduced, the beam exhibited significant expansion in the horizontal direction, particularly in the pre-target region. This expansion was attributed to increased momentum spread ($\Delta p/p$) resulting from the shortened bunch length. \\
The narrowest bunch length achieved was 3.0 ns at the medium beam intensity, corresponding to a proton beam sigma of approximately 750 ps. However, further reduction in the bunch length led to challenges in beam stability, necessitating careful tuning of the Booster parameters.

\section{Results of Bunch Rotation Studies}
This study focused on optimizing the beam in the Booster to achieve a narrow bunch length while managing the resulting challenges, such as increased energy spread and dispersion. By varying intensity and extraction time, we successfully shortened the bunch length at both high and medium intensities. The impact of collimators on beam losses and intensity was evaluated, showing that collimators, while slightly increasing losses, also enhanced beam intensity. The analysis highlighted the importance of collimation, particularly in managing losses along the BNB line. The implementation pathway includes exploring additional collimation strategies and utilizing medium beam intensity to achieve the desired beam profile with minimal losses.
Overall, the study demonstrated the ability to achieve sub-nanosecond bunch lengths, which are crucial for improving the precision of the stroboscopic approach in neutrino experiments.

\subsection{Beam diagnostic system}
In this study, we utilize two primary beam diagnostic systems. 
The first is the Beam Loss Monitor (BLM) to assess beam loss along both the Booster Ring and the 8 GeV beam transport beam line. 
The second is the Resistive Wall Current Monitor (RWM) to measure the beam current and the beam bunch length. 
A brief description of each beam diagnostic system is provided below. 

\subsubsection{Beam loss monitor}
The BLM is a coaxial ionization chamber. 
Beam losses during acceleration were systematically investigated and well understood \cite{kapin2017booster} based on BLM data. 
In this study, we confirmed that no significant beam loss occurs during acceleration, even when the longitudinal phase space is rotated in the Booster ring. However, there was an outstanding beam loss in the 8 GeV beam transport line. 
A detail description of the beam loss observation is provided in the following section. 

\subsubsection{Wall current monitor}
The Resistive Wall Current Monitor (RWM) \cite{bunchlength2016} measures the mirror charge current of the proton beam in both the Booster ring and the 8 GeV beam transport line. 
It provides both the beam current and the beam bunch length. 
However, due to the fast nature of the signal, cable dispersion can introduce additional challenges.
In this analysis, we did not apply any corrections for signal dispersion. 
Instead, a Fast Fourier Transform (FFT) analysis was used to filter out signal noise. 

\begin{figure}[htbp]
    \centering
    \includegraphics[width=0.45\textwidth]{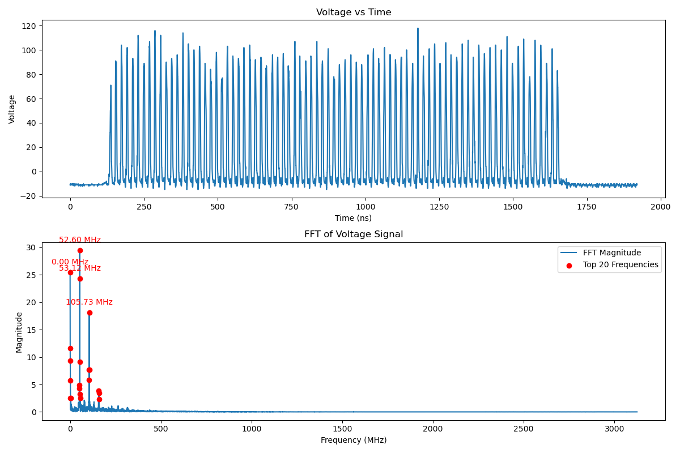} 
    \caption{Top plot shows the raw signal from the RWM. 
    Bottom plot shows the output from FFT.}
    \label{fig:FFT}
\end{figure}

Figure \ref{fig:FFT} shows the raw signal from the RWM and output from FFT. 
The frequency domain plot shows that a primary frequency is around 53 MHz. 
This corresponds to the RF frequency. 
The sencond order of the RF frequency is also observed. 
Other frequency components can be a background noise. 

\begin{figure}[htbp]
    \centering
    \begin{subfigure}[t]{0.5\textwidth}
        \centering
        \includegraphics[width=0.45\textwidth]{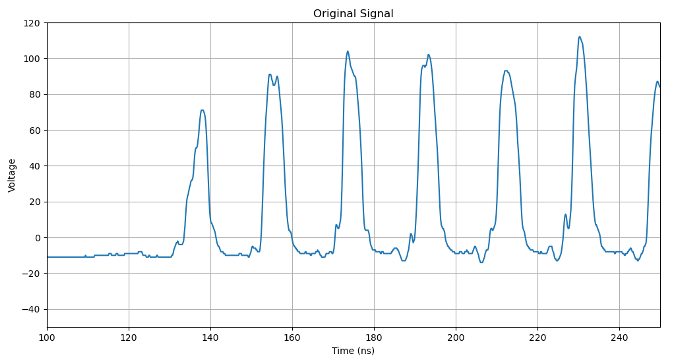}
    \end{subfigure}
    \hspace{0.02\textwidth} 
    \begin{subfigure}[t]{0.5\textwidth}
        \centering
        \includegraphics[width=0.45\textwidth]{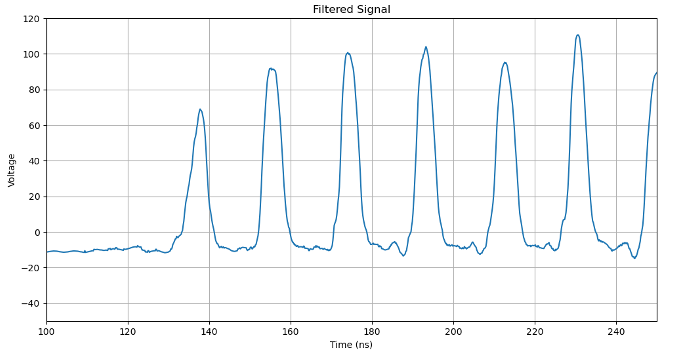} 
    \end{subfigure}
    \caption{The enlarged raw RWM signal and after FFT filtering.}
    \label{fig:FFT_filter}
\end{figure}

Figure~\ref{fig:FFT_filter} shows that the raw RWM signal and after the FFT signal subtraction. 
The RWM signal becomes smooth after the FFT filtering. 
The RMS of beam bunch length is estimated using the FFT filtered RWM data with a Gaussian fitting. 
Consequently, the difference of estimated RMS between with and without filtering is quite similar. 
Therefore, we estimate the bunch length by using the raw data in the rest of the document. 

\subsection{Beam Loss Characterization}
In the study, we explored tuning the beam in the Booster to achieve the narrowest possible bunch length by varying extraction time, intensity, and frequency. The beam was sent through the BNB beamline to SBND and ANNIE. Our investigation included managing undesirable side effects such as the increased energy spread resulting from bunch shortening, which led to increased dispersion in the transverse horizontal direction.

\begin{figure}[htbp]
    \centering
    \includegraphics[width=0.51\textwidth, trim=0 0 0 140, clip]{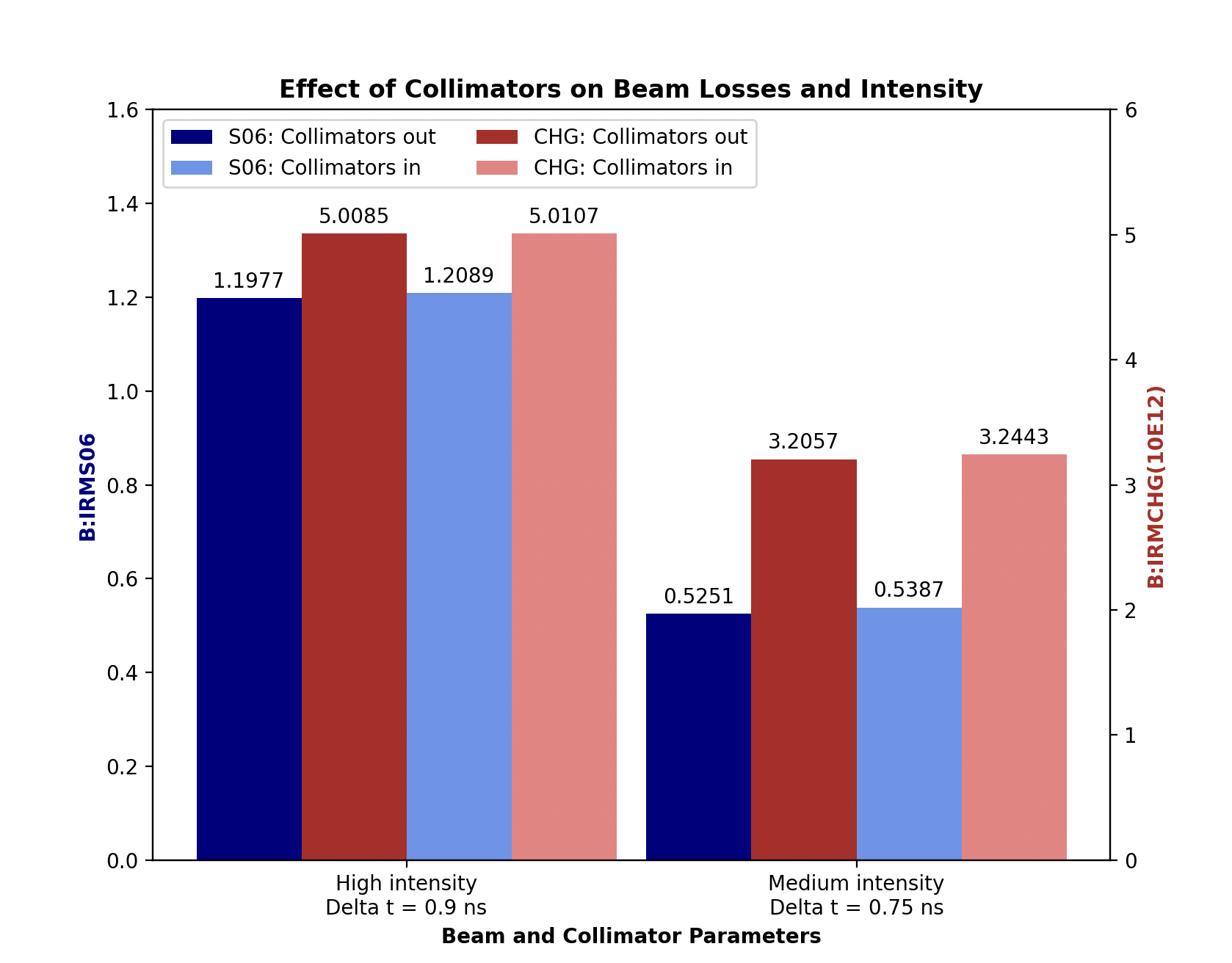}
    \caption{Effect of collimators on beam losses and intensity at different bunch lengths and intensities. The plot shows the comparison between high and medium intensity settings, with and without collimators.}
    \label{fig:collimator_effect}
\end{figure}

Results at high intensity showed that the nominal bunch length of 2.1 ns could be shortened to 0.9 ns. At medium intensity, the bunch length was reduced from 1.725 ns to 0.75 ns. The study examined the effect of collimators on beam losses and intensity (Figure \ref{fig:collimator_effect}). While inserting collimators slightly increases beam losses, it also boosts beam intensity.

\begin{figure}[htbp]
    \centering
    \includegraphics[width=0.51\textwidth, trim=0 0 0 41.5, clip]{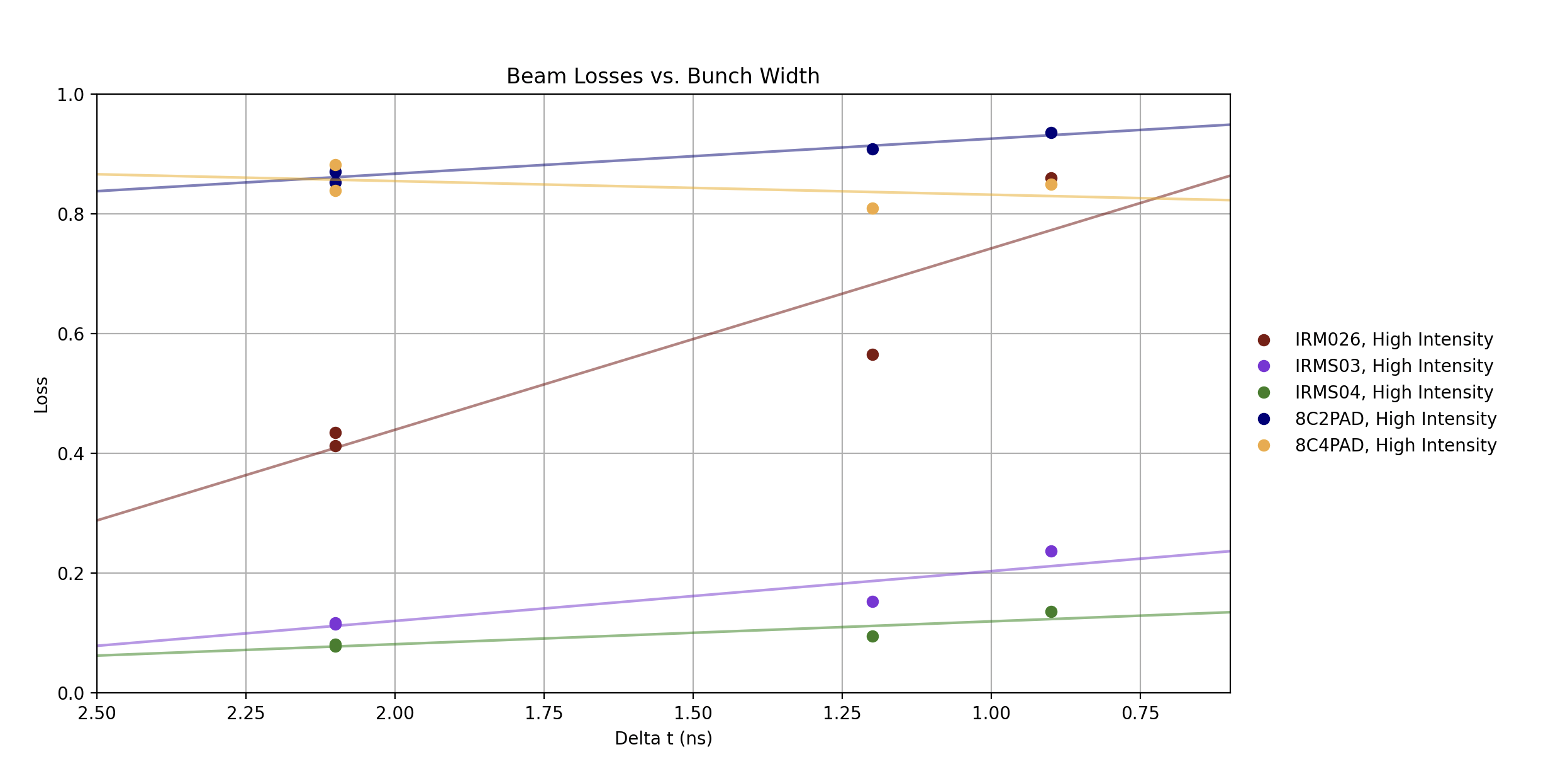}
    \caption{Beam losses vs. bunch length ($\Delta$t) as recorded by various beam loss monitors (BLMs) and intensity monitors (IRMs) in the Booster. The plot shows a non-linear increase in losses as bunch lengths shorten. IRM026, IRM503, IRM504, and the loss monitors near the 8 GeV collimator (8C2PAD, 8C4PAD) are all represented. Standard error did not exceed 0.01 at any point. Losses were calculated by averaging raw data over several minutes, with some exclusions made for irregularities.}
    \label{fig:beam_losses_vs_bunch_width}
\end{figure}

Figure \ref{fig:beam_losses_vs_bunch_width} illustrates the relationship between beam losses and bunch length ($\Delta$t) as measured by various beam loss monitors (BLMs) and intensity monitors (IRMs) in the Booster. As the bunch lengths were shortened from 2.5 ns to 0.75 ns, beam losses increased non-linearly across nearly all BLMs, including IRM026, IRM503, IRM504, and the loss monitors near the 8 GeV collimator (8C2PAD, 8C4PAD). Despite this increase in losses, the standard error did not exceed 0.01 at any of these points. The loss data was averaged over several minutes, with some exclusions made for irregularities.

\begin{figure}[htbp]
    \centering
    \begin{subfigure}[t]{0.5\textwidth}
        \centering
        \includegraphics[width=0.45\textwidth]{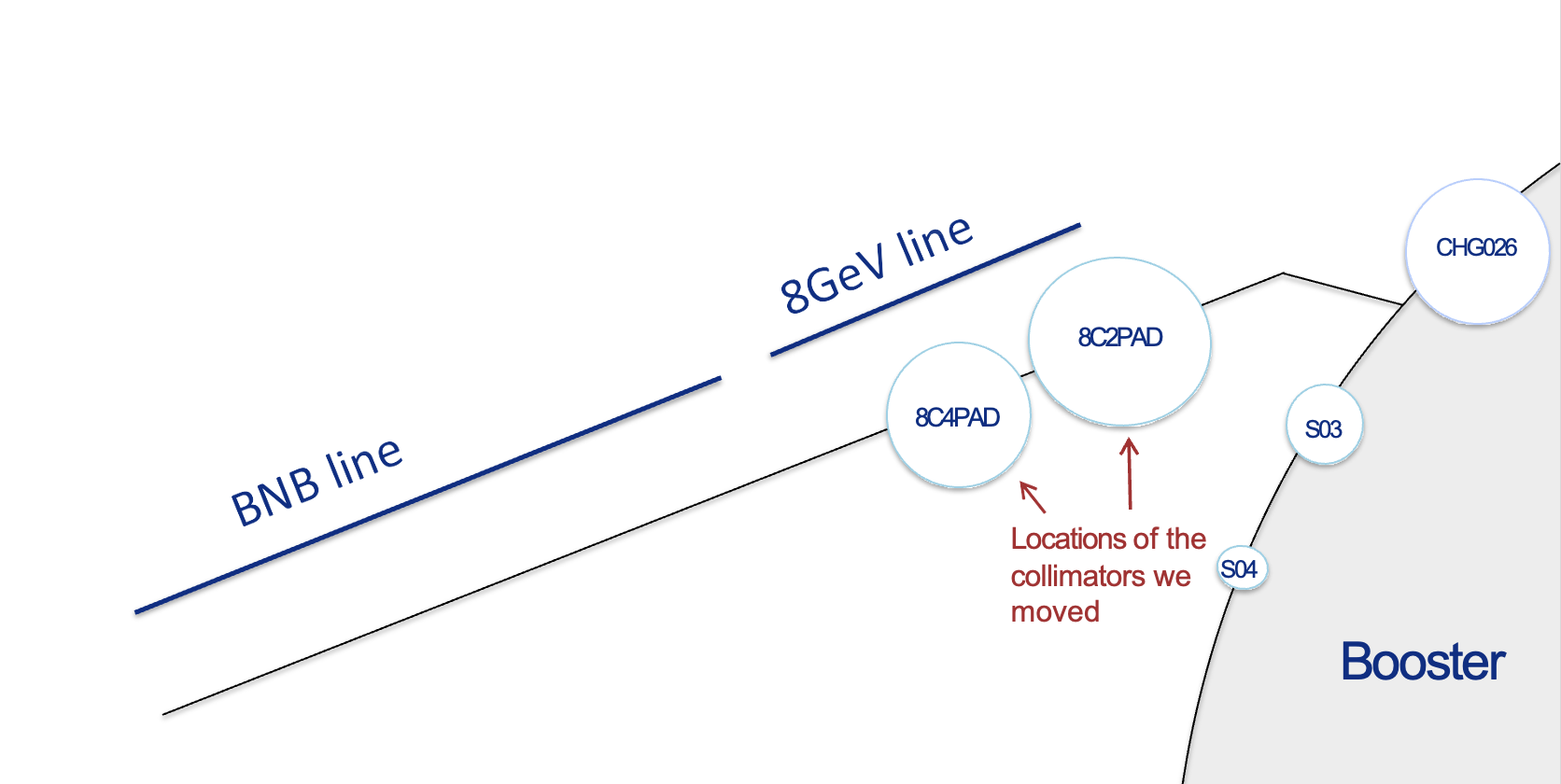}
    \end{subfigure}
    \hspace{0.02\textwidth} 
    \begin{subfigure}[t]{0.5\textwidth}
        \centering
        \includegraphics[width=0.45\textwidth]{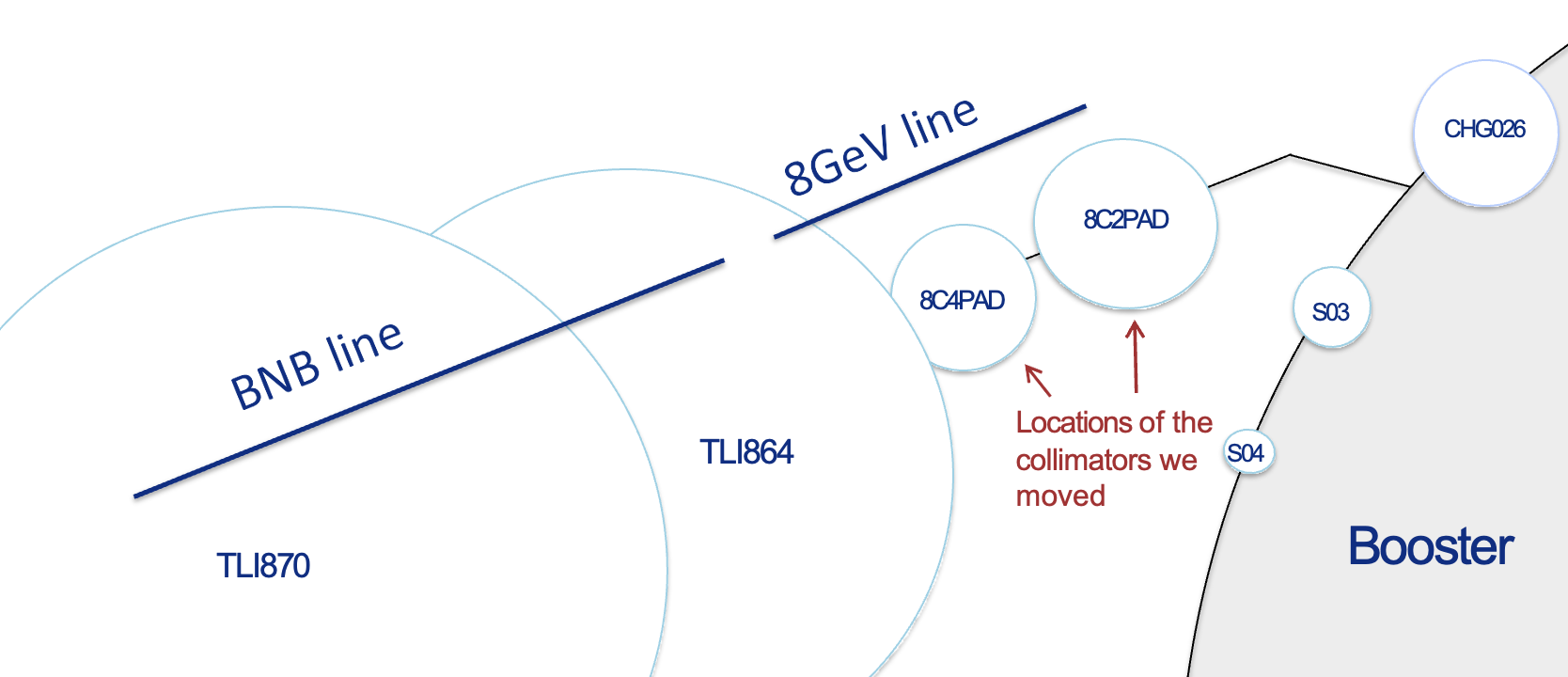} 
    \end{subfigure}
    \caption{Cartoons showing the distribution of beam losses by location when collimators are removed. Losses were greatest near extraction sites and in the BNB line, particularly at monitors in the 8 GeV line (8C2PAD, 8C4PAD) and near CHG026. The bottom cartoon shows additional locations and their associated beam losses along the BNB line, with circles indicating the relative magnitude of losses.}
    \label{fig:collimator_losses_by_location_combined}
\end{figure}

Next, we examine the distribution of beam losses by location for the high-intensity case with the collimators removed. The losses increased significantly in some locations, particularly near extraction sites and along the BNB line. The cartoon in Figure \ref{fig:collimator_losses_by_location_combined} shows the approximate locations of key beam loss monitors (BLMs) relative to the beamline. Notably, losses were greatest at monitors positioned in the 8 GeV line (8C2PAD, 8C4PAD) and near the CHG026 and S03/S04 locations in the Booster. The circles in the cartoon give a rough idea of the magnitude of losses relative to their location along the beamline, with larger circles indicating higher losses. The locations marked with arrows indicate where collimators were moved to manage these losses. The BNB losses are particularly concerning, as highlighted by the large circles on the diagram. This visualization underscores the need for effective collimation strategies to mitigate these troubling losses.

\begin{figure}[htbp]
    \centering
    \includegraphics[width=0.45\textwidth]{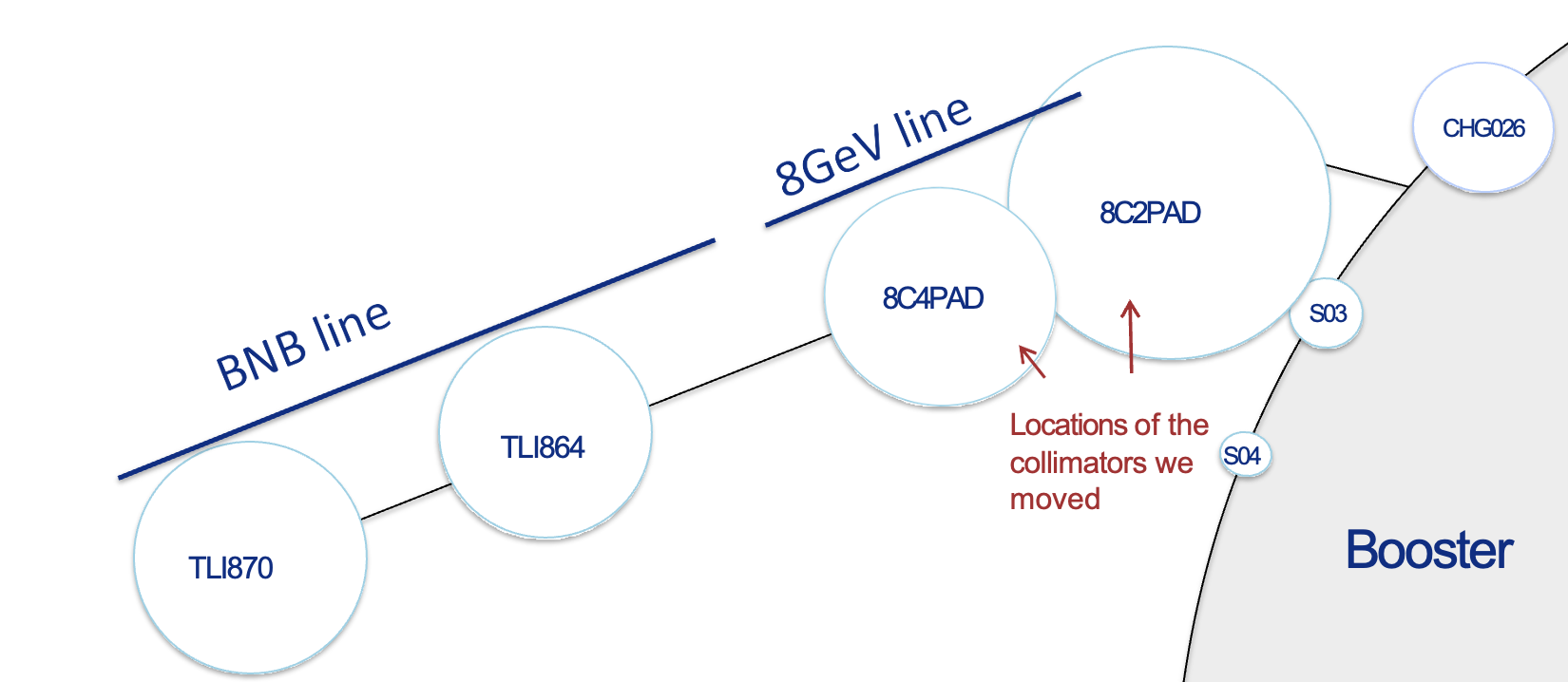} 
    \caption{Distribution of beam losses by location with collimators inserted. The plot shows a significant reduction in losses along the BNB line, a smaller reduction in Booster losses, and a slight increase in the 8 GeV line. Overall, losses were well below the accepted maximums when running at the medium intensity mode.}
    \label{fig:collimator_losses_in}
\end{figure}

Figure \ref{fig:collimator_losses_in} illustrates the distribution of beam losses by location when running in medium intensity mode with collimators inserted. The cartoon shows a significant reduction in losses along the BNB line, with a smaller reduction in the Booster and a slight increase in the 8 GeV line, all within acceptable limits.

\begin{figure}[htbp]
    \centering
    \includegraphics[width=0.51\textwidth, trim=0 0 0 10, clip]{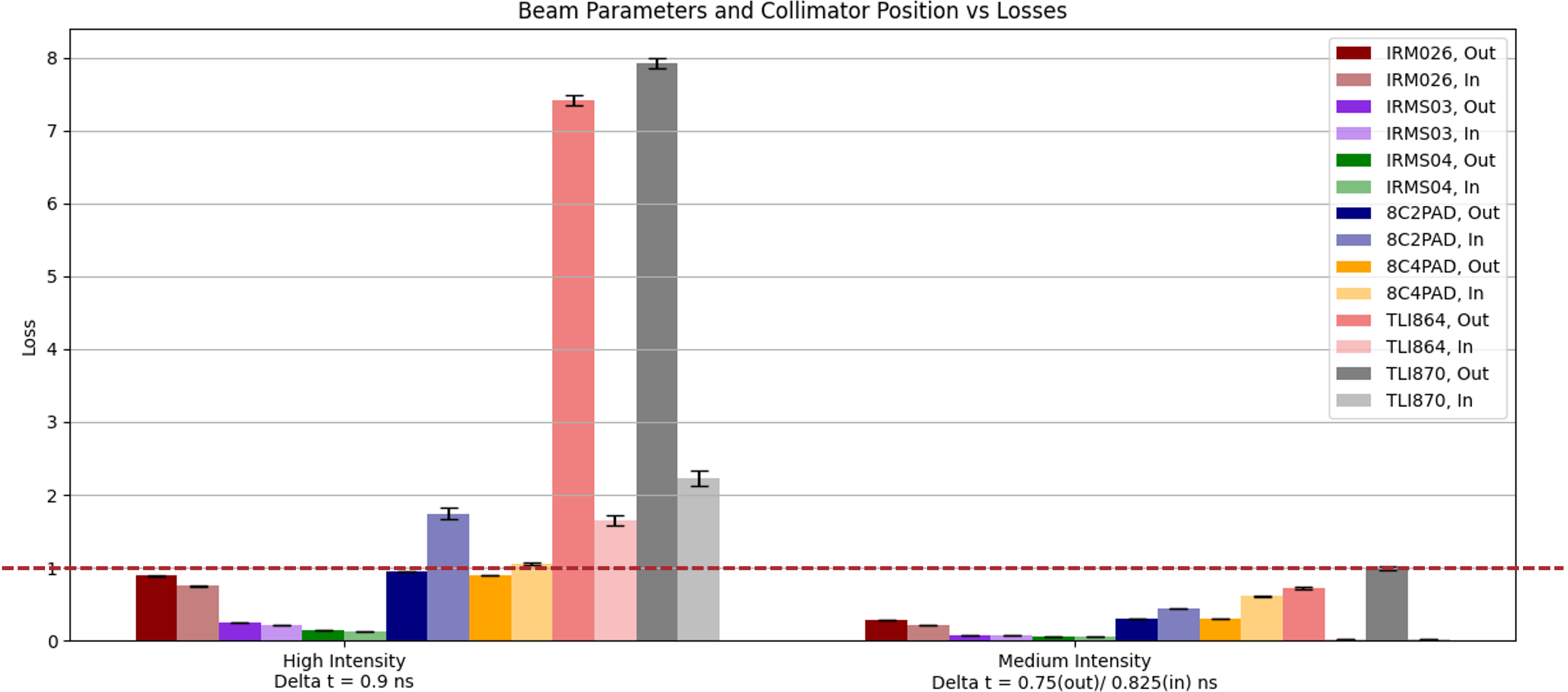} 
    \caption{Effect of collimation on beam losses. Pink and gray bars represent losses in the BNB line. The results indicate that collimation is crucial for managing beam losses, especially in high-intensity operations with narrow bunch lengths. The maximum acceptable loss threshold is marked by the dashed red line.}
    \label{fig:collimation_effect}
\end{figure}

Next, we examine the effect of inserting collimators on the distribution of beam losses by location. As illustrated in Figure \ref{fig:collimator_losses_in}, there was a significant reduction in losses along the BNB line, while the Booster saw a smaller reduction, and there was a slight increase in losses along the 8 GeV line. Importantly, the losses remained well below the accepted maximums when operating at medium intensity.\\
Figure \ref{fig:collimation_effect} illustrates the effect of collimation on beam losses across different sections of the beam line. The various colored bars represent data from loss monitors placed at key locations along the beam path, such as:

- \textbf{IRM026} (extraction point from the 8 GeV line)
- \textbf{IRM053 and IRM054} (within the Booster)
- \textbf{8C2PAD and 8C4PAD} (near the 8 GeV collimators)
- \textbf{TLI864 and TLI870} (in the BNB beamline)

The bars differentiate between the conditions with collimators \textbf{out} (darker bars) and \textbf{in} (lighter bars) for both high- and medium-intensity beam settings. For \textbf{high-intensity beams} ($\Delta t$ = 0.9 ns), significant losses are recorded in the BNB line when collimators are out, as seen by the large pink and gray bars. These losses drastically decrease when collimators are inserted, highlighting the critical role of collimation in minimizing losses, especially at high intensities.

In \textbf{medium-intensity settings} ($\Delta t$ = 0.75/0.825 ns), the collimators also help reduce beam losses, though the effect is less dramatic compared to high-intensity settings. The monitors near the BNB collimators (TLI864 and TLI870) show the most noticeable reduction in losses when collimators are engaged.

The dashed red line represents the maximum acceptable beam loss threshold. The data indicates that with collimation, the beam losses remain well within safe operational limits, particularly in the BNB line, where losses are critical to maintaining beam stability and minimizing damage.

\begin{figure}[htbp]
    \centering
    \includegraphics[width=0.51\textwidth, trim=0 0 0 10, clip]{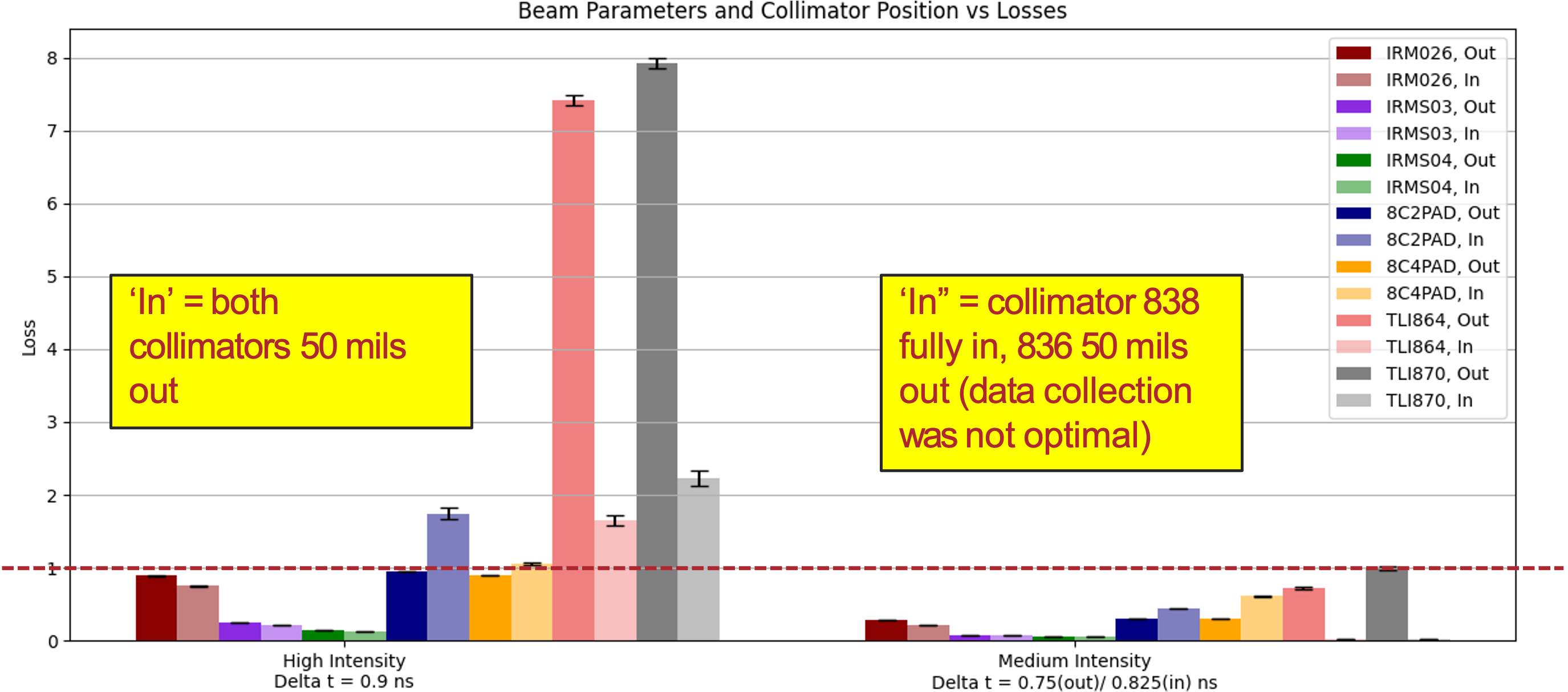} 
    \caption{Effect of collimation on beam losses at different intensity levels. The plot shows that at high intensity, partially inserted collimators (150 mils in) were stable, but fully inserting collimator 838 caused the beam to trip. Losses at moderate intensity were acceptable with collimators inserted, as indicated by the bars staying below the maximum acceptable loss threshold (dashed red line).}
    \label{fig:collimation_effect_tripping}
\end{figure}

At high intensity, we observed (shown in Figure \ref{fig:collimation_effect_tripping}) that running with collimators 836/838 partially inserted (150 mils in) was stable, but the beam tripped when we attempted to fully insert collimator 838. The exact cause of the tripping remains unidentified and requires further investigation. On the other hand, no tripping was observed at moderate intensity, and losses were significantly lower, particularly with the collimators inserted.

\begin{figure}[htbp]
    \centering
    \includegraphics[width=0.21\textwidth, trim=0 0 0 10, clip]{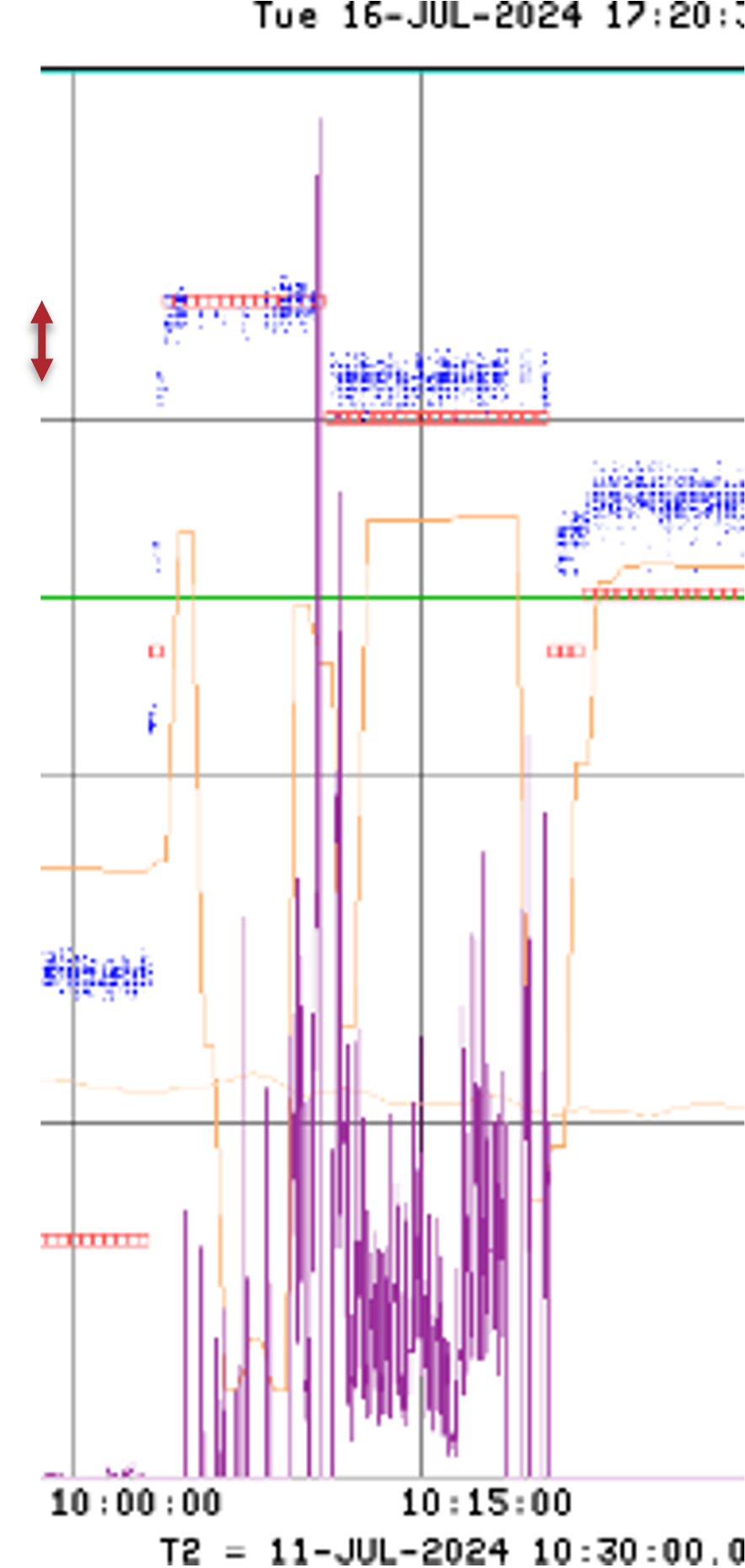} 
    \caption{Effect of a slight reduction in intensity on BNB losses. The plot shows that reducing the intensity from 4.7E12 to 4.5E12 resulted in a significant decrease in BNB losses, as indicated by the TLI864 data (purple). Intensity data is shown in blue.}
    \label{fig:BNB_losses}
\end{figure}

It was noted that a slight reduction in intensity from 4.7E12 to 4.5E12 led to a significant decrease in BNB losses, as seen in the TLI864 data (shown in purple in Figure \ref{fig:BNB_losses}). The vertical arrow on the plot indicates the small reduction in intensity, which corresponds to the drop from 4.7E12 to 4.5E12. 
Preliminary results suggest that even small adjustments in intensity can have a substantial impact on reducing losses. The intensity data is depicted in blue, and the accompanying plot highlights the correlation between the slight drop in intensity and the observed reduction in losses.

The transverse beam profiles were also measured near the target after the beam was narrowed longitudinally.
    \begin{figure}[htbp]
    \centering
    \includegraphics[width=0.51\textwidth]{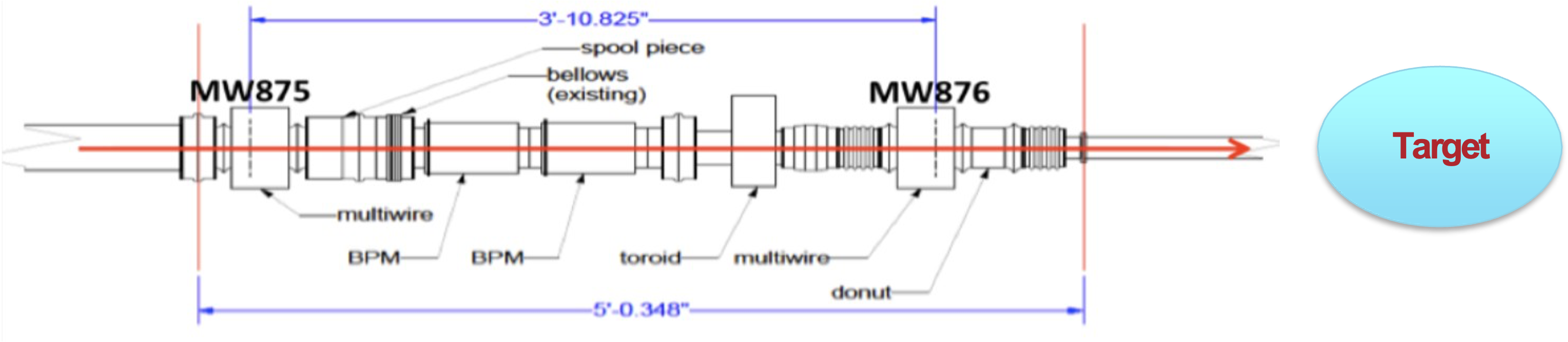} 
    \caption{Diagram showing the locations of MW875 and MW876 upstream of the target. These devices are crucial for accurately measuring the transverse beam profile as it approaches the target. Data from MW876, in particular, reflects the beam profile directly as it impinges on the target, providing essential information for beam performance assessment.}
    \label{fig:MW876_diagram}
\end{figure}

We focused on MW875 and MW876, positioned just upstream of the target as shown in Figure \ref{fig:MW876_diagram}. The data from MW876 should provide a precise reflection of the beam's profile as it reaches the target, which is vital for understanding and optimizing the performance of the narrow beam. 

\begin{figure}[htbp]
    \centering
    \includegraphics[width=0.45\textwidth]{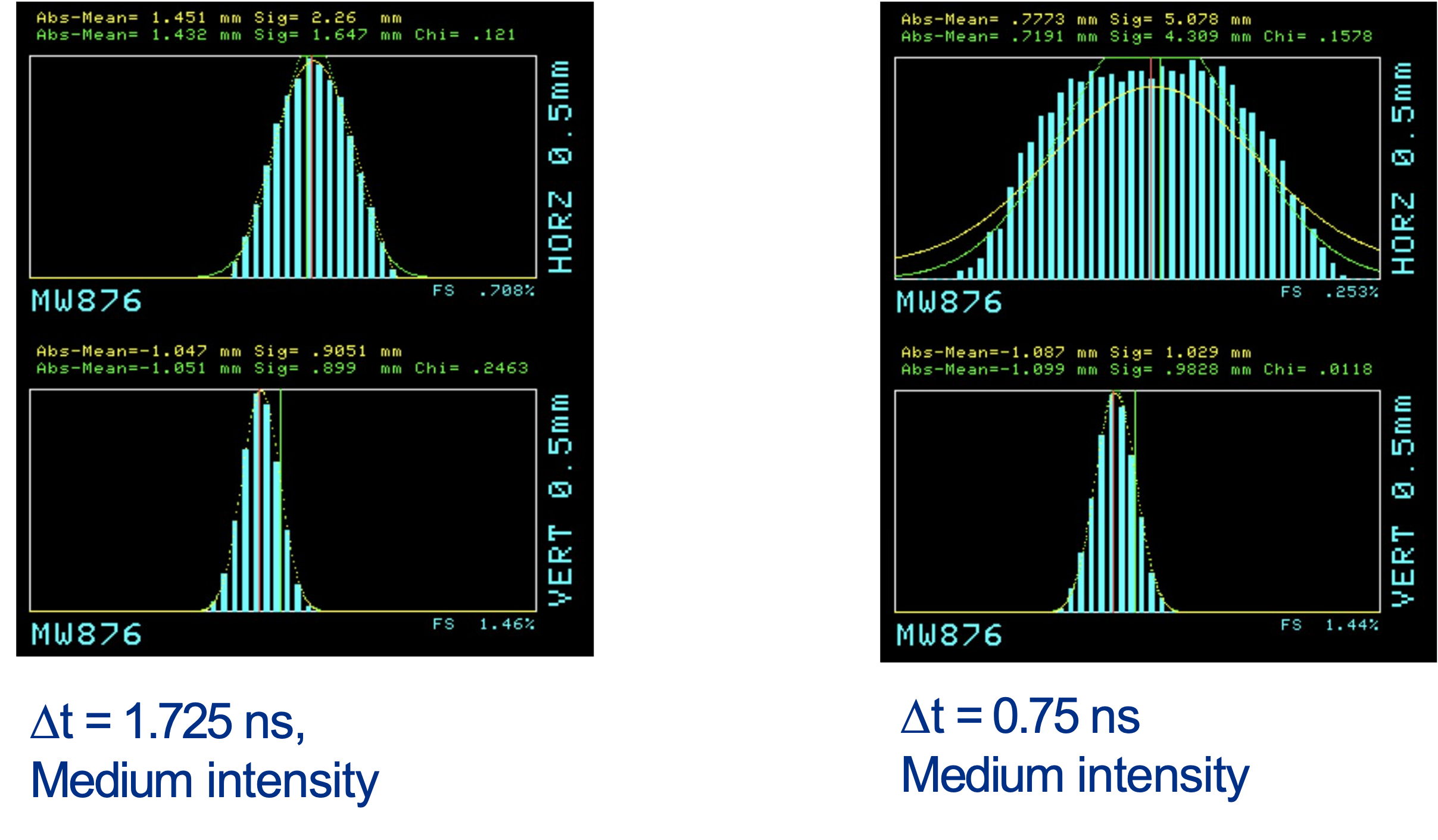} 
    \caption{Transverse beam profiles at medium intensity for two different bunch lengths. The left profile corresponds to a bunch length of $\Delta t = 1.725$ ns, displaying a more Gaussian shape. The right profile shows the beam at a shorter bunch length of $\Delta t = 0.75$ ns, where the spot size increased, and the distribution became irregular and non-Gaussian.}
    \label{fig:transverse_beam_profile}
\end{figure}

\begin{figure}[htbp]
    \centering
    \includegraphics[width=0.45\textwidth]{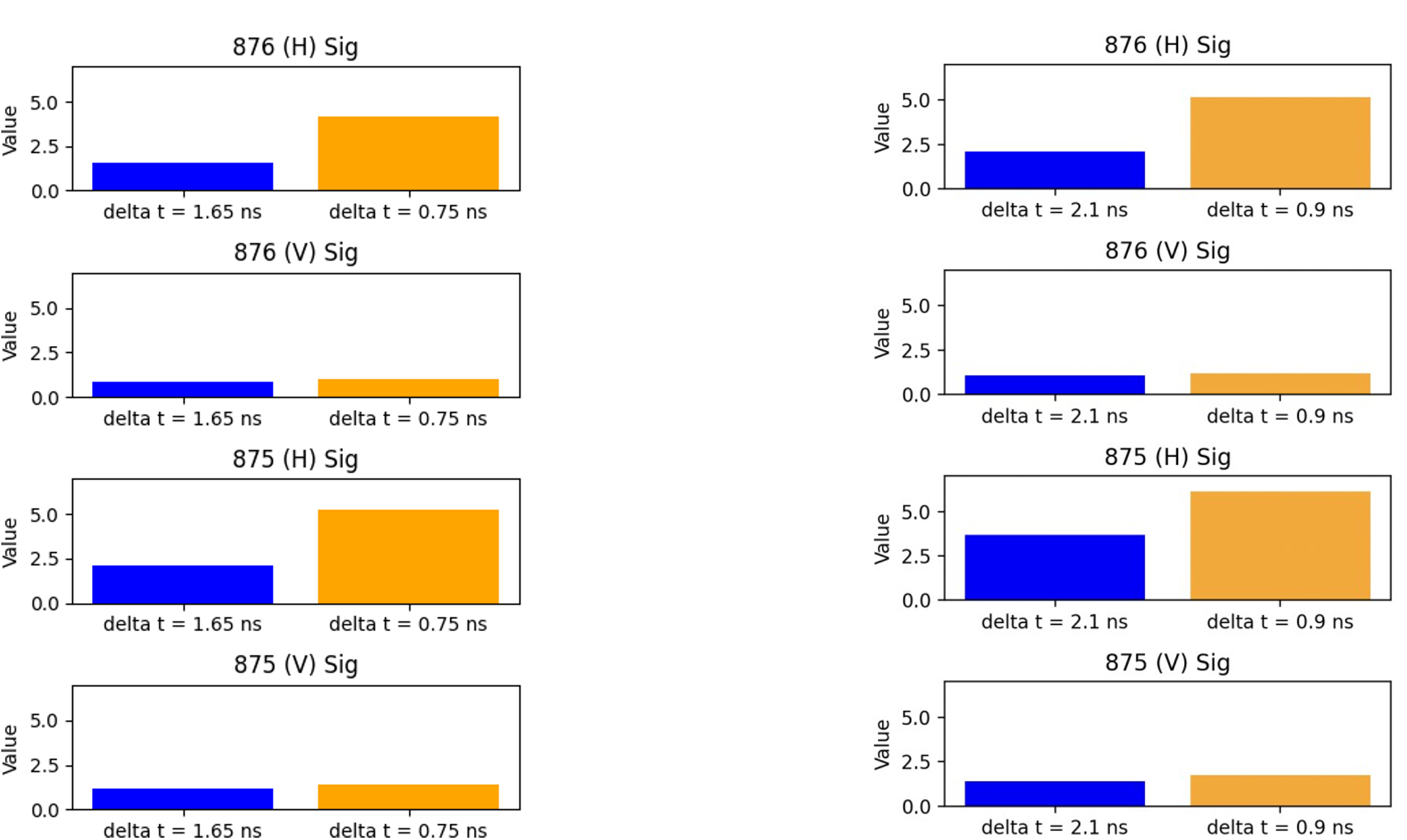} 
    \caption{Changes in the transverse beam profile before and after bunch shortening. The left column shows the profile changes at an intensity of 2.8E12, and the right column at 4.7E12. The horizontal (H) and vertical (V) signals from monitors MW875 and MW876 highlight how the cross-section becomes more asymmetrical as the bunch length is reduced.}
    \label{fig:transverse_beam_profile_changes}
\end{figure}

The beam's transverse profile was analyzed at two different bunch lengths, both at medium intensity as shown in Figure \ref{fig:transverse_beam_profile}. The left plot shows the beam profile with a longer bunch length of 1.725 ns, where the distribution maintains a relatively Gaussian shape. However, when the bunch length was shortened to 0.75 ns (as shown on the right), the spot size increased, and the distribution became irregular and non-Gaussian.
The transverse beam profile became asymmetrical after the bunch length was shortened. The plots in Figure \ref{fig:transverse_beam_profile_changes} show how the beam's shape changes between monitors MW875 and MW876 for both horizontal (H) and vertical (V) signals. The left plots correspond to the medium intensity, and the right plots correspond to the high intensity mode. In both cases, when the bunch length was reduced, the beam's shape became less regular and more uneven.

\begin{figure}[htbp]
    \centering
    \includegraphics[width=0.41\textwidth]{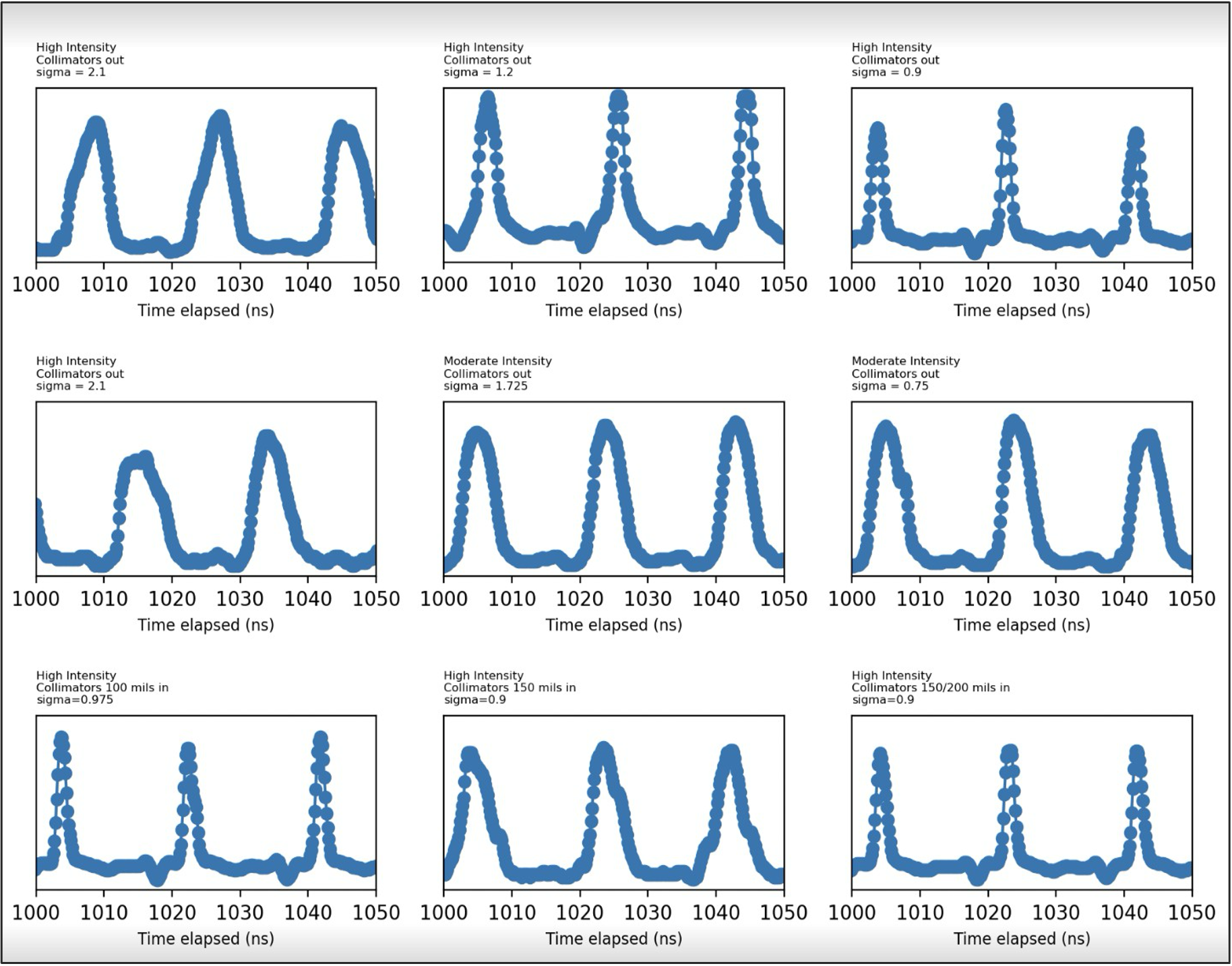} 
    \caption{Longitudinal beam profile generated from RWM data, showing a generally Gaussian, unimodal, and symmetric shape. Differences in the appearance of the troughs can be observed, which may be influenced by the bunch length and intensity. Note that the y-axes are scaled arbitrarily.}
    \label{fig:longitudinal_beam_profile}
\end{figure}

The horizontal beam profile becomes more uneven after bunch length narrowing primarily due to the increased dispersion in the horizontal plane of the Booster Neutrino Beam (BNB) line. Dispersion causes particles with slightly different momenta to spread out more in the horizontal direction as they travel through the beamline. When the bunch length is shortened, this dispersion effect becomes more pronounced, leading to greater irregularities and a more uneven horizontal profile. The vertical profile, with less dispersion, remains relatively unaffected.
The longitudinal beam profile, as generated from the RWM data and shown in Figure \ref{fig:longitudinal_beam_profile}, generally remains Gaussian, unimodal, and symmetric, indicating a stable beam shape even after changes in beam conditions. However, there is a noticeable difference in the appearance of the troughs in the profile depending on the bunch length and intensity. This variation in the troughs could indicate slight distortions in the beam structure as the bunch length is shortened or the intensity is altered. The y-axes of the plots have been scaled arbitrarily.

\begin{figure}[htbp]
    \centering
    \begin{subfigure}[t]{0.5\textwidth}
        \centering
        \includegraphics[width=0.41\textwidth, trim=0 0 0 50, clip]{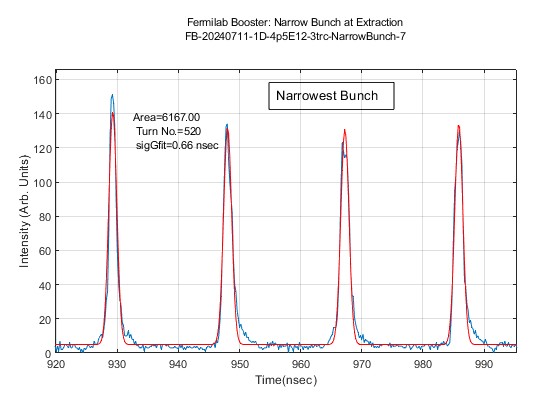}
    \end{subfigure}
    \hspace{0.05\textwidth} 
    \begin{subfigure}[t]{0.5\textwidth}
        \centering
        \includegraphics[width=0.41\textwidth, trim=10 10 10 2.12, clip]{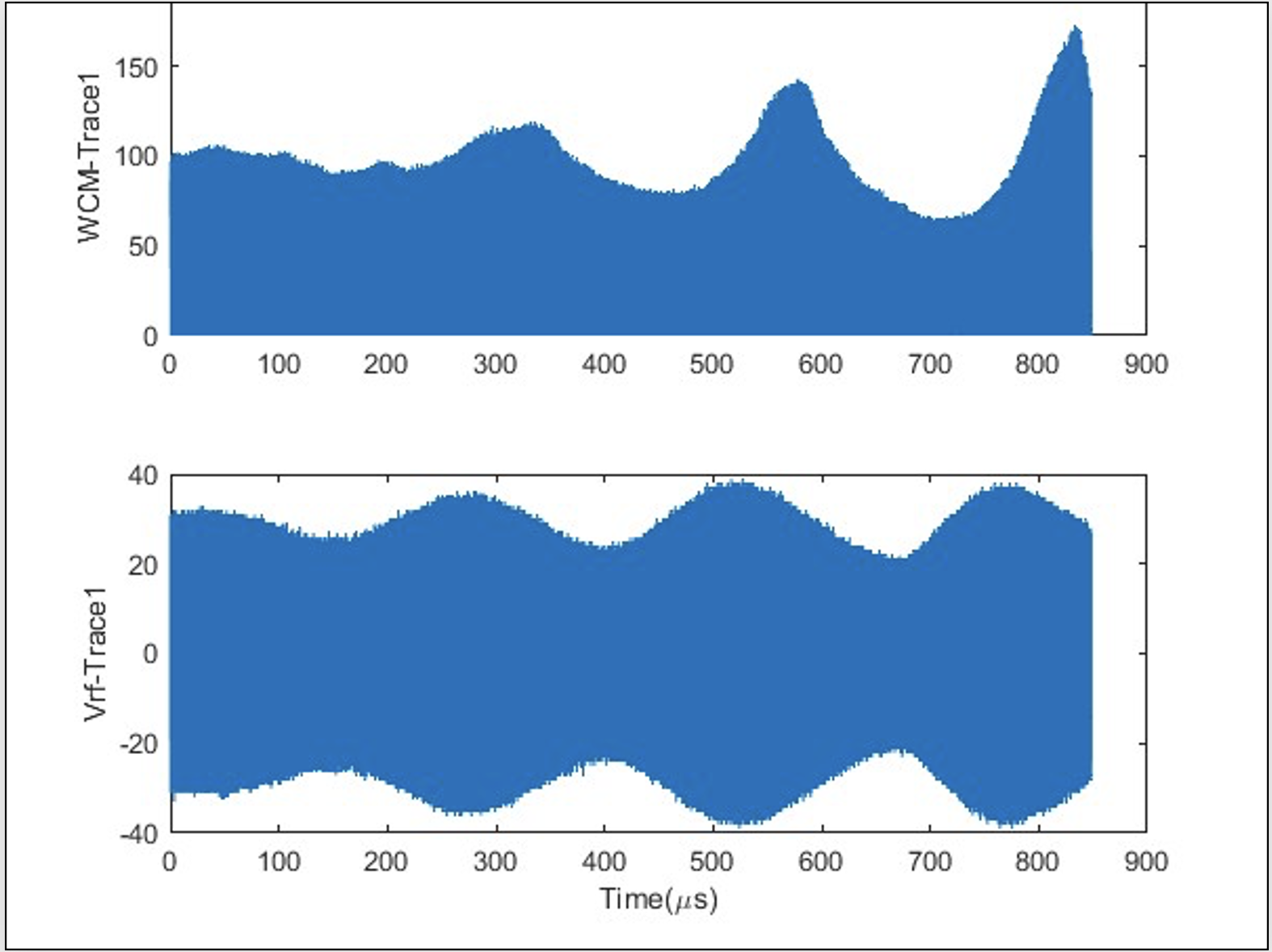} 
    \end{subfigure}
    \caption{Narrow bunch extraction data from Fermilab Booster. The top plot shows the intensity profile of the narrowest bunch, while the bottom plot shows the Wall Current Monitor (WCM) and RF traces.}
    \label{fig:narrow_bunch_data}
\end{figure}

A further attempt to narrow the beam in high-intensity mode was made by fine-tuning the extraction time. Figure \ref{fig:narrow_bunch_data} shows the intensity profile of the narrowest bunch at extraction (turn No. 520), with an RMS width of 0.66 ns. This represents the narrowest bunch width achieved during this data-taking attempt in high-intensity mode. The wall current monitor (WCM) data shows how the beam evolves over time as it undergoes bunch rotation. The peaks in the plot indicate the bunching of particles, and the increasing amplitude towards the right suggests that the bunching is becoming more pronounced closer to extraction. The RF trace provides the phase information of the beam at various time slices.

\subsection{Implementation Pathway}
To achieve the desired beam profile and minimize losses in the narrow booster beam scheme, we will focus on several key steps. We will explore additional collimation strategies to reduce the beam spot size while minimizing losses, particularly by collimating the booster beam before extraction. Operating at medium beam intensity could strike a balance between reducing bunch length and controlling losses. This approach will involve iterative testing to optimize beam intensity, collimation, and loss management. Additionally, the upcoming addition of two new collimators in the 8 GeV line may enable more efficient operation at higher intensity in the narrow bunch beam mode, potentially improving overall performance.

\section{BNB Optics Analysis}
\begin{figure}[htbp]
    \centering
    \includegraphics[width=0.40\textwidth]{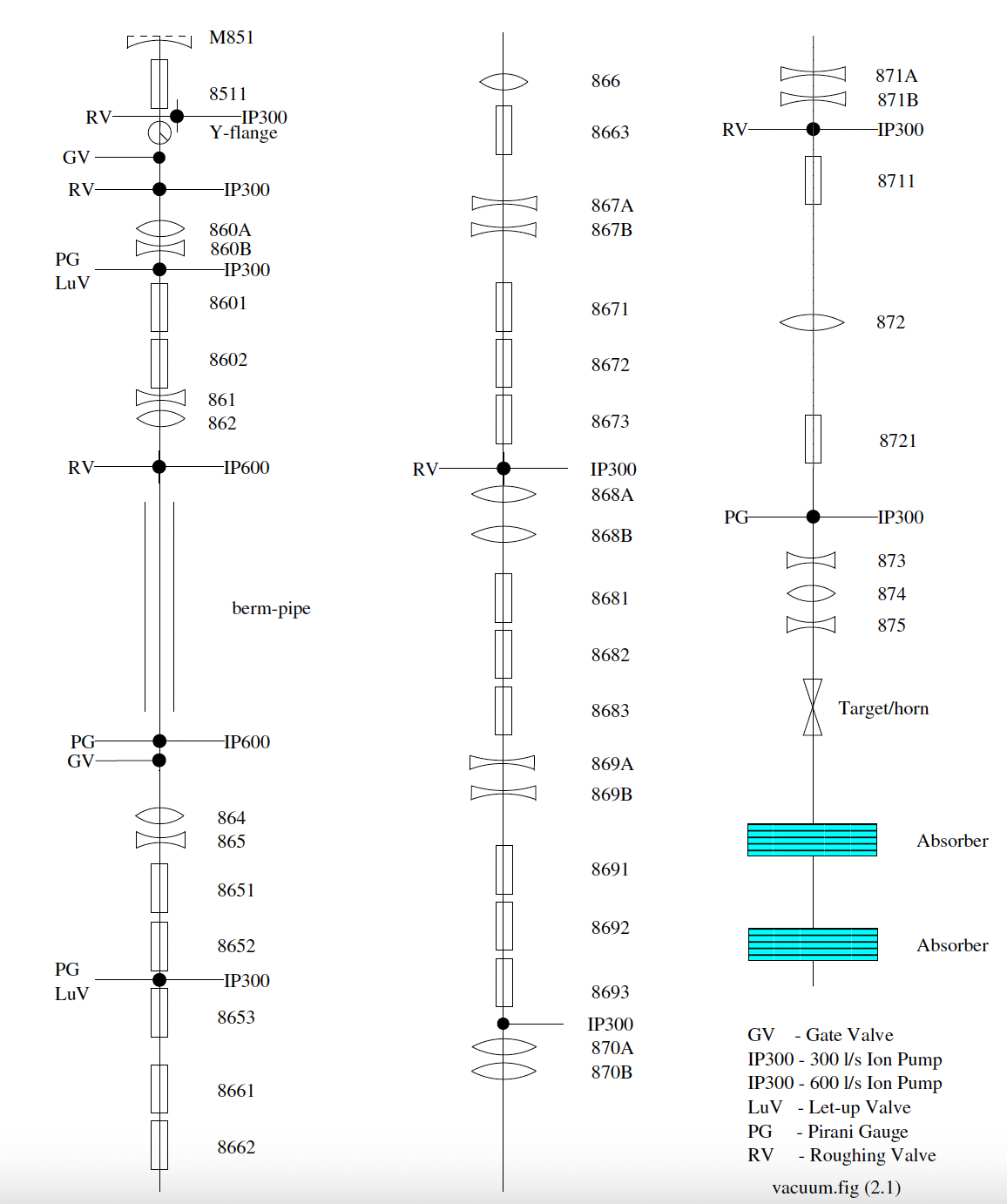} 
    \caption{Map of the BNB beamline, highlighting key locations such as D876 where dispersion plays a critical role in beam dynamics.}
    \label{fig:Beamline_Map}
\end{figure}

\begin{figure}[htbp]
    \centering
    \includegraphics[width=0.41\textwidth]{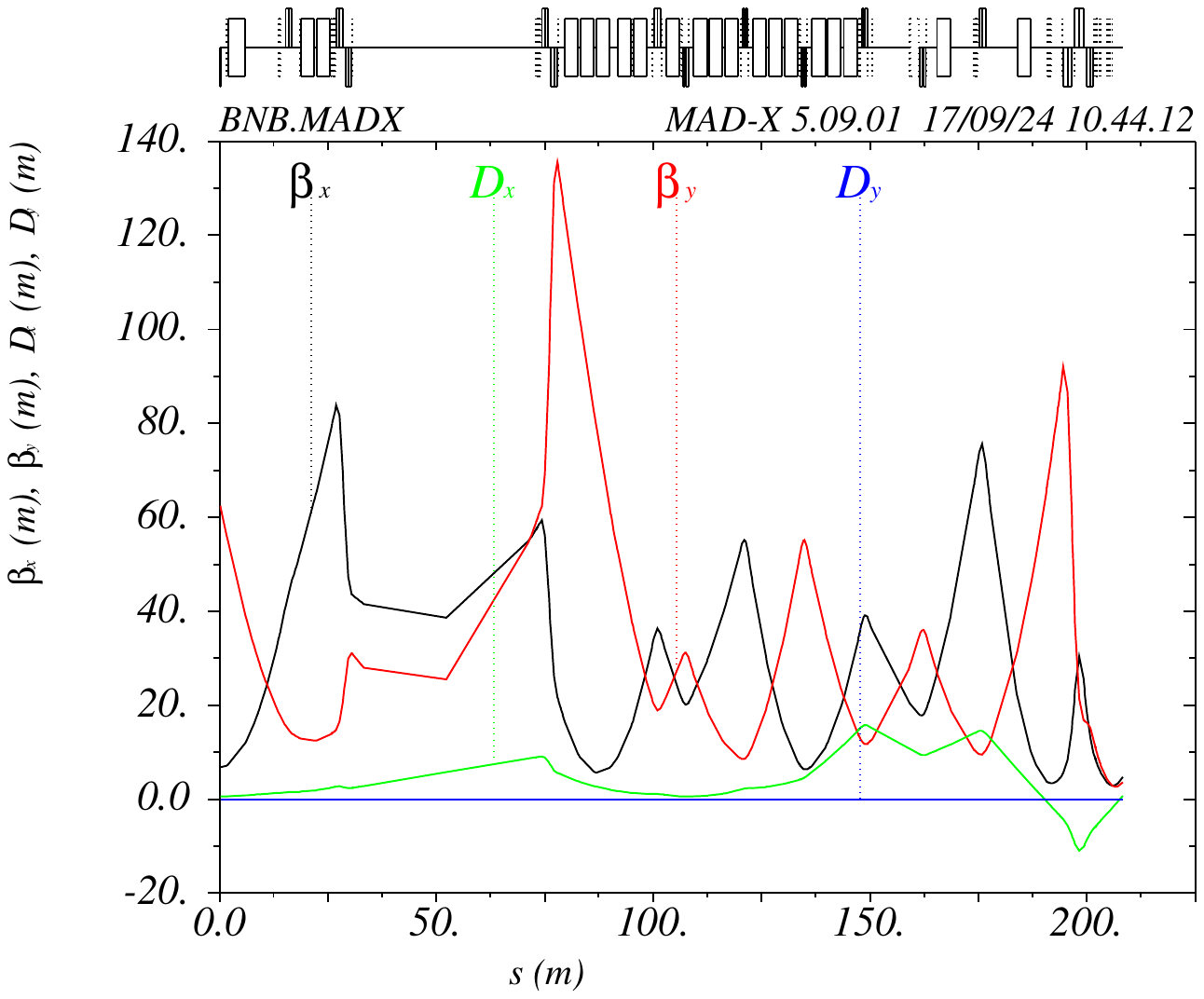} 
    \caption{Preliminary BNB MADX study showing the $\beta$ function and the dispersion function $D_x$ and $D_y$ at critical points along the beamline. The spot size $\sigma_x$ at location D876 is influenced by the dispersion and is used to calculate the momentum spread.}
    \label{fig:Dispersion_Plot}
\end{figure}

Our immediate focus is on investigating potential modifications to the beam optics, along with a comprehensive analysis of MW profile data at additional locations along the beamline. This effort includes evaluating the dispersion function at critical points, within the BNB, and examining pre-target dispersion to reduce any excess energy spread.
 The map of the beamline (Figure \ref{fig:Beamline_Map}) \cite{boone_tdr} provides an overview of key locations of the BNB beamline. 
The dispersion plot (Figure \ref{fig:Dispersion_Plot}) generated from the MADX \cite{madx} study illustrates how dispersion varies along the beamline.

At the pre-target location, we estimated the momentum spread at D876, situated between location 875 and the target as shown in Figure \ref{fig:Beamline_Map}).
The relative momentum spread $\delta$ was calculated using:

\[
\delta = \frac{p - p_0}{p_0}
\]

Where:
\begin{itemize}
    \item $\delta$ is the momentum spread
    \item $p$ is the momentum at the location of interest
    \item $p_0$ is the reference momentum
\end{itemize}
$p$ is the momentum at D876 and $p_0$ is the reference momentum. $\delta$ is the deviation from the reference momentum $p_0$. It quantifies how much the momentum $p$ at a given point (in this case, D876) differs from the reference momentum.

The spot size $\sigma_{x,\text{rms}}$ is influenced by the beta function $\beta_x$, the RMS emittance $\epsilon_{\text{rms}}$, and the dispersion $D$. The relationship is given by:

\[
\sigma_{x,\text{rms}}^2 = \beta_x \epsilon_{\text{rms}} + D^2 \delta^2
\]

For the calculation of the momentum spread at D876, we used the RMS spot size $\sigma_{x,\text{rms}}^2$ value of $2.5788084 \times 10^{-5}$ m$^2$, which was derived from the 1$\sigma$ spot size provided in the TDR. The beta function ($\beta_x$) and the dispersion ($D$) at D876 were obtained from the MADX calculation, with values of 5.834905169 meters and -4.706790002 meters, respectively. These inputs were then used to determine the momentum spread, $\delta$, and the resulting momentum difference from the reference momentum, $p_0$.

Using these given values for $\sigma_{x,\text{rms}}$, $\beta_x$, $\epsilon_{\text{rms}}$, and $D$ at D876, the momentum difference $(p - p_0)$ was calculated to be approximately 6.85 MeV/c.

\section{Conclusions and Recommendations}
Neutrino beams will play a crucial role in future neutrino experiments. Using the stroboscopic approach, neutrino beams can be exploited to their fullest potential.
The study demonstrated that by fine-tuning the extraction time, beam intensity, and RF settings, it is possible to achieve sub-nanosecond bunch lengths. However, these adjustments must be made cautiously to avoid excessive dispersion and beam instability, which can compromise the quality of the experiment.
The next steps for achieving efficient operation of the BNB beam in narrow bunch mode, will involve carefully balancing beam intensity and collimation strategies. We will explore running the beam at moderate intensity with fully inserted collimators to manage losses while maintaining a narrow bunch profile. Additionally, a detailed analysis of post-target data and potential modifications to the beam optics will be conducted, focusing on mitigating dispersion effects along the beamline and in pre-target areas. These efforts are crucial to optimizing the beam conditions for narrow bunch mode, ensuring high efficiency and data quality in upcoming experimental runs.

\end{document}